\documentclass[twocolumn,apj]
{openjournal}
\usepackage{amssymb}

\usepackage{color,comment}
\usepackage[normalem]{ulem}
\usepackage{newtxtext,newtxmath}
\usepackage{amsmath} 
\usepackage{ctable}
\usepackage[breaklinks,colorlinks,citecolor=blue,urlcolor=blue]{hyperref}

\newcommand{\orcidauthor}[3]{\author{\href{http://orcid.org/#1}{#2$^{#3}$}}}

\newcommand{\qpah}[1]{$q_{\rm PAH}$ }
\newcommand{\lpah}[1]{$L_{\rm PAH}$ }
\newcommand{\lfir}[1]{$L_{\rm FIR}$ }

\shorttitle{}
\shortauthors{Narayanan et al.}

\begin{document}

\title[]{The Growth of Dust in Galaxies in the First Billion Years with Applications to Blue Monsters\vspace{-1.25cm}}

\orcidauthor{0000-0002-7064-4309}{Desika Narayanan}{1,2*}
\orcidauthor{0000-0002-5653-0786}{Paul Torrey}{3,4,5}
\orcidauthor{0000-0001-6106-5172}{Daniel P. Stark}{6}
\orcidauthor{0000-0002-0302-2577}{John Chisholm}{7,8}
\orcidauthor{0000-0001-8519-1130}{Steven L. Finkelstein}{7,8}
\orcidauthor{0000-0002-8111-9884}{Alex M. Garcia}{3,4,5}
\orcidauthor{0000-0003-3816-7028}{Federico Marinacci}{9,10}
\orcidauthor{0000-0002-7727-1824}{Jessica Kelley-Derzon}{11}
\orcidauthor{0000-0002-3790-720X}{Laura V. Sales}{12}
\orcidauthor{0000-0002-2919-1109}{Ethan Savitch}{1}
\orcidauthor{0000-0001-8593-7692}{Mark Vogelsberger}{13,14}
\orcidauthor{0009-0008-7017-5742}{Dhruv T. Zimmerman}{1}
\affiliation{$^{1}$Department of Astronomy, University of Florida, 211 Bryant Space Sciences Center, Gainesville, FL 32611 USA}
\affiliation{$^{2}$Cosmic Dawn Center at the Niels Bohr Institute, University of Copenhagen and DTU-Space, Technical University of Denmark}
\affiliation{$^{3}$Department of Astronomy, University of Virginia, 530 McCormick Road, Charlottesville, VA 22903, USA}
\affiliation{$^{4}$Virginia Institute for Theoretical Astronomy, University of Virginia, Charlottesville, VA 22904, USA}
\affiliation{$^{5}$The NSF-Simons AI Institute for Cosmic Origins, USA}
\affiliation{$^{6}$Department of Astronomy, University of California, 501 Campbell Hall \#3411, Berkeley, CA 94720, USA}
\affiliation{$^{7}$Department of Astronomy, The University of Texas at Austin, 2515 Speedway, Stop C1400, Austin, TX 78712, USA}
\affiliation{$^{8}$Cosmic Frontier Center, The University of Texas at Austin, Austin, TX 78712, USA}
\affiliation{$^{9}$Department of Physics and Astronomy "Augusto Righi", University of Bologna, via Gobetti 93/2, 40129, Bologna, Italy}
\affiliation{$^{10}$INAF, Astrophysics and Space Science Observatory Bologna, Via P. Gobetti 93/3, 40129 Bologna, Italy}
\affiliation{$^{11}$Department of Physics, University of Florida, 2001 Museum Road, Gainesville, FL 32611 USA}
\affiliation{$^{12}$Department of Physics and Astronomy, University of California, Riverside, CA 92507, USA}
\affiliation{$^{13}$Department of Physics and Kavli Institute for Astrophysics and Space Research, Massachusetts Institute of Technology, Cambridge, MA 02139, USA}
\affiliation{$^{14}$Fachbereich Physik, Philipps Universit\"at Marburg, D-35032 Marburg, Germany}

\thanks{$^*$E-mail: \href{mailto:desika.narayanan@ufl.edu}{desika.narayanan@ufl.edu}}



\begin{abstract}
  A combination of JWST observations at $z \approx 12-14$ and ALMA
  observations of extremely dust-rich systems at $z \approx 6$ has
  demonstrated that dust grows extremely fast in the early Universe,
  with galaxies amassing up to $10^7 M_\odot$ of dust in just $500$
  Myr between $z=12\rightarrow6$.  In this paper we demonstrate, via a series of numerical
  experiments conducted in cosmological zoom-in simulations, that a
  likely pathway for this dust accumulation in the first formed
  galaxies is through production at early times via supernovae, followed
  by the rapid growth on ultrasmall dust grains.  Our main results
  follow.  The stellar production of dust dominates until $z \sim
  10-11$ at which point galaxies transition to a growth-dominated
  regime.  We employ a Shapley analysis to demonstrate that the local
  density is the dominant factor driving dust growth, followed by the
  grain size distribution.  A rapid rise in the small-to-large grain
  ratio with decreasing redshift (owing to grain-grain shattering)
  drives growth through increased dust surface area per unit mass.  Growth
  models are necessary to match the dust content of ALMA detected
  sources at $z \sim 6$.  Finally, we demonstrate that ``blue
  monsters'', massive, UV-bright galaxies at $z>10$ with extremely
  blue continuum slopes likely have dust-to-stellar mass
  ratios $10^{-4}-10^{-3}$, but their top-heavy grain size distributions render them
  optically thin in the UV, providing a natural explanation for their
  observed properties without requiring exotic dust geometries.

\end{abstract}

\keywords{Galaxies, Galaxy Formation, Dust formation, Interstellar Medium, JWST}

\section{Introduction}
In its first few years of operations, JWST has discovered a large
number of galaxies at $z>10$ characterized both by extremely bright UV
luminosities, as well as low inferred interstellar dust masses \citep{arrabalharo23a,curtislake23a,austin23a,
bunker24a,cullen24a,morales24a,roberts-borsani24a,topping24a}.  The
inference of low dust masses comes from two different methodologies,
both of which lead to similar results.  The first of these
methodologies is through measurements of the UV continuum power law
slope in these early galaxies.  Formally, the UV spectral energy
distribution (SED) near $\lambda = 1500 \AA$ is often described as a
power law \citep[$F_\lambda \propto \lambda^\beta$;][]{calzetti94a}.
This power law index (or sometimes colloquially referred to as a ``UV
slope'') is primarily impacted by dust reddening
\citep{calzetti97a,meurer99a,bouwens10a,bouwens12a,finkelstein12a,dunlop12a,narayanan18a},
although older stellar populations
\citep{reddy18a,calabro21a,tacchella22a}, and nebular emission
\citep[e.g.][]{byler17a,katz24a,topping22a,topping24a,cullen24a} can
also contribute to reddening in a secondary manner
\citep{narayanan25a}.  Although the specific expected $\beta$ for a truly
dust-free population varies depending on the model, most studies have
narrowed the expected dust-free UV slope to $\beta_0 \approx -2.7-3$
for low-metallicity systems
\citep{bouwens10a,raiter10a,wilkins12a,bouwens16a,stanway16a,jaacks18a,reddy18a,narayanan25a}.
Indeed, a number of rest frame UV observations of $z>10$ galaxies have
evidenced extremely blue $\beta$ slopes, consistent with negligible
dust reddening
\citep[e.g.][]{austin24a,cullen24a,morales24a,topping24a}.

The second method for constraining the dust mass in UV-bright $z>10$
galaxies is through SED fitting techniques.  In practice, the visual
attenuation ($A_{\rm V}$) is determined, that with an estimate of the
effective radius, can be converted into a dust mass
\citep{ferrara24b,ferrara25a}. SED fitting techniques carry some
uncertainty owing to the age-reddening-metallicity degeneracy, which is
further complicated by the potential impact of bursty star formation 
histories and outshining on early Universe inference
\citep{tacchella22a,topping22a,whitler23a,ciesla24a,narayanan24a,mosleh25a,wang25a}
-- though see \citet{cochrane25a} and \citet{harvey25a} for
counter-examples. Nevertheless, the general results from these techniques 
agree with those from $\beta$ measurements. Quantitatively, massive galaxies ($M_* \approx
10^8-10^9 M_\odot$) at $z>10$ are constrained to have $\lesssim 5 \times
10^4 M_\odot$ of dust, far below the typical dust-to-stellar mass
ratio for lower redshift star-forming galaxies
\citep[e.g.][]{watson15a,bouwens16a,marrone18a,fudamoto21a,endsley22a,sommovigo22a,ferrara25a}.

At the same time, there is a clear boundary condition: numerous ALMA
detections of galaxies with stellar masses $M_* \approx 10^9-10^{10}
M_\odot$ at $z\sim6$ have evidenced copious dust reservoirs, with dust
masses $\sim 10^7-10^8 M_\odot$
\citep[e.g.][]{watson15a,knudsen17a,laporte17a,marrone18a,hashimoto19a,bakx21a,fudamoto21a,endsley22a,ferrara22a,topping22a,sommovigo22a,algera24a}.
These detections, coupled with the inference of relatively dust-poor
galaxies only a few hundred Myr earlier, imply extremely rapid dust
growth in the early Universe.  Specifically, the combination of JWST
and ALMA constraints on dust masses during the Epoch of Reionization suggest
a growth of $\sim 10^7 M_\odot$ of dust in just $500-600$ Myr for galaxies $M_* \approx 10^{9}-10^{11} M_\odot$ at $z\approx6$.

These two sets of observations -- of relatively dust-free galaxies at
$z>10$, and fairly dusty systems at $z\sim6$, immediately leads to two
significant problems in the early Universe.  The first problem is: how
do galaxies attain such significant dust reservoirs over such a short
time frame?

Model predictions fall into two generic camps for the rapid rise of
dust in the early Universe: (i) {\it Production Dominated} and (ii)
{\it Growth Dominated}.  Production-dominated schemes suggest that
Type II Supernovae (SNe), which dominate dust production in the early
Universe, can generate enough dust to satisfy the observational
constraints at lower ($z \approx 6$) redshifts if they produce $\sim 0.5-1.5 M_\odot$ of dust per supernova
\citep[e.g.][]{triani20a,dayal22a}.    To produce $\sim 1 M_\odot$ of dust in a type II supernova requires 
a $100\%$ condensation efficiency of all refractory elements in a 
massive $25 M_\odot$ explosion\footnote{The mass
fraction of dust that survives a supernova is still fairly uncertain
\citep{todini01a,bianchi07a,cherchneff10a,matsuura11a,barlow10a,delooze17a}, with observations suggesting more than $1$ order of magnitude uncertainty.} \citep{woosley95a,nomoto06a,dwek07a}. While it is a
heavy burden on SNe to produce the massive dust reservoirs observed by
$z \sim 6$, \citet{michalowski15a} compute that this may still be a
feasible mechanism for explaining the $z \sim 6$ dust masses, so long as 
there is effectively no loss in dust mass owing to dust
destruction in SNe blast waves \citep[see, e.g.][]{shchekinov25a}. 
However, if the dust masses at $z\sim 5-7$  are underestimated, as may be
expected for a reasonable distribution of dust temperatures in
galaxies \citep{sommovigo25a}, then this only further exacerbates the
problem\footnote{It is worth noting that at least some models suggest that the dust masses at high-redshift may be {\it over}-estimated \citep{choban24a}.}.

As a result of the aforementioned challenges associated with
production-only models, a second school of thought has emerged that
dust {\it growth} in the ISM is necessary to satisfy the $z \approx 6$
dust mass constraints.  Observations of galaxy dust masses at a range
of redshifts ($z=0-6$) show clear trends between the dust to metal
mass ratio as a function of metallicity, as well as the related dust
to gas ratio as a function of metallicity, suggesting the growth of
dust as a function of the available metal mass
\citep{magdis12a,remyruyer14a,devis19a,heintz23a,popping22a,popping23a,algera24a,algera25a,burgarella25a}.
Accordingly, while individual details may differ, simulations using a
diverse range of both semi-analytic and hydrodynamic modeling schemes
find that accretion is the dominant means for growing dust masses in
early Universe galaxies
\citep{popping17a,li19a,vijayan19a,graziani20a,esmerian22a,parente22a,dicesare23a,esmerian23a,lewis23a,lower23a,parente23a,lower24a,choban25a,narayanan25a}. 

These competing pictures -- between production based models and a
litany of growth based models -- allow us to sharpen our framing of
the first major question: {\it When} does growth take over from
production in dominating the dust budget, and what physical processes
drive the growth of dust in the early Universe?

The second major question that has emerged in the study of early
Universe dusty galaxies is that of the origin of ``blue monsters'' --
a class of massive, UV-bright ($M_{\rm UV} \la -20-21$) and yet extremely
blue ($\beta \la -2.2$; therefore, seemingly not dusty) galaxies at
$z>10$ \citep{ziparo23a,ferrara25a}.  A key result from production
dominated first dust scenarios (which we will additionally demonstrate
with our simulations in this paper) is that galaxies typically have dust to
stellar mass ratios ($M_{\rm dust}/M_*$) $\sim 10^{-4}$ from production alone 
\citep[e.g.][]{dayal22a,ferrara25a}.  Growth-dominated models tend to
increase this ratio by $\sim 1-2$ orders of magnitude, $M_{\rm dust}/M_*
\sim 10^{-3}$ \citep{narayanan25a}.  This said, while the inference of
stellar masses of blue monsters is that they are already quite massive
by $z\sim10$, with inferred $M_* \sim 10^8-10^9 M_\odot$, their
extremely blue UV continuum slopes imply inferred dust to stellar mass
ratios $\lesssim 10^{-5}$ \cite{ferrara23a,ferrara24a}.

  These order(s) of magnitude discrepancies between the inferred and
  expected dust to stellar mass ratios in blue monsters has inspired a
  class of models known as ``attenuation free models'' \citep[AFM;
  ][]{ferrara23a,ziparo23a,ferrara25a}, in which the dust was formed
  -- as expected -- but launched into an optically thin geometry owing
  to early stellar winds \citep{ferrara24a}.  This said, there is as
  of yet no observational confirmation of the AFM model
  \citep{ferrara25b}, and the origin of blue monsters remains an
  unsolved puzzle in the physics of early Universe galaxy formation.

  The goal of this paper is to present a series of numerical
  experiments, using cosmological zoom-in simulations of galaxy
  evolution, to both understand the driving physics behind the growth
  of the first dust in the Universe, as well as the origin of the
  extremely blue and massive blue monster population at $z>10$.  In
  \S~\ref{section:methods}, we discuss our numerical methodology. In
  \S~\ref{section:physical_evolution}, we lay the foundation for the
  physical properties of the first galaxies within our
  simulation framework.  In \S~\ref{section:dust_growth}, we use this
  shared understanding of the evolution of early galaxy properties to
  develop a model for the rapid rise in galaxy dust content from
  $z=14\rightarrow7$, and in \S~\ref{section:blue_monsters}, tie this
  model to the origin of Blue Monsters.  In
  \S~\ref{section:discussion}, we discuss our findings in the context
  of alternative theories for early Universe dust, and in
  \S~\ref{section:summary}, we summarize.


  \begin{figure*}

    \includegraphics[scale=0.4]{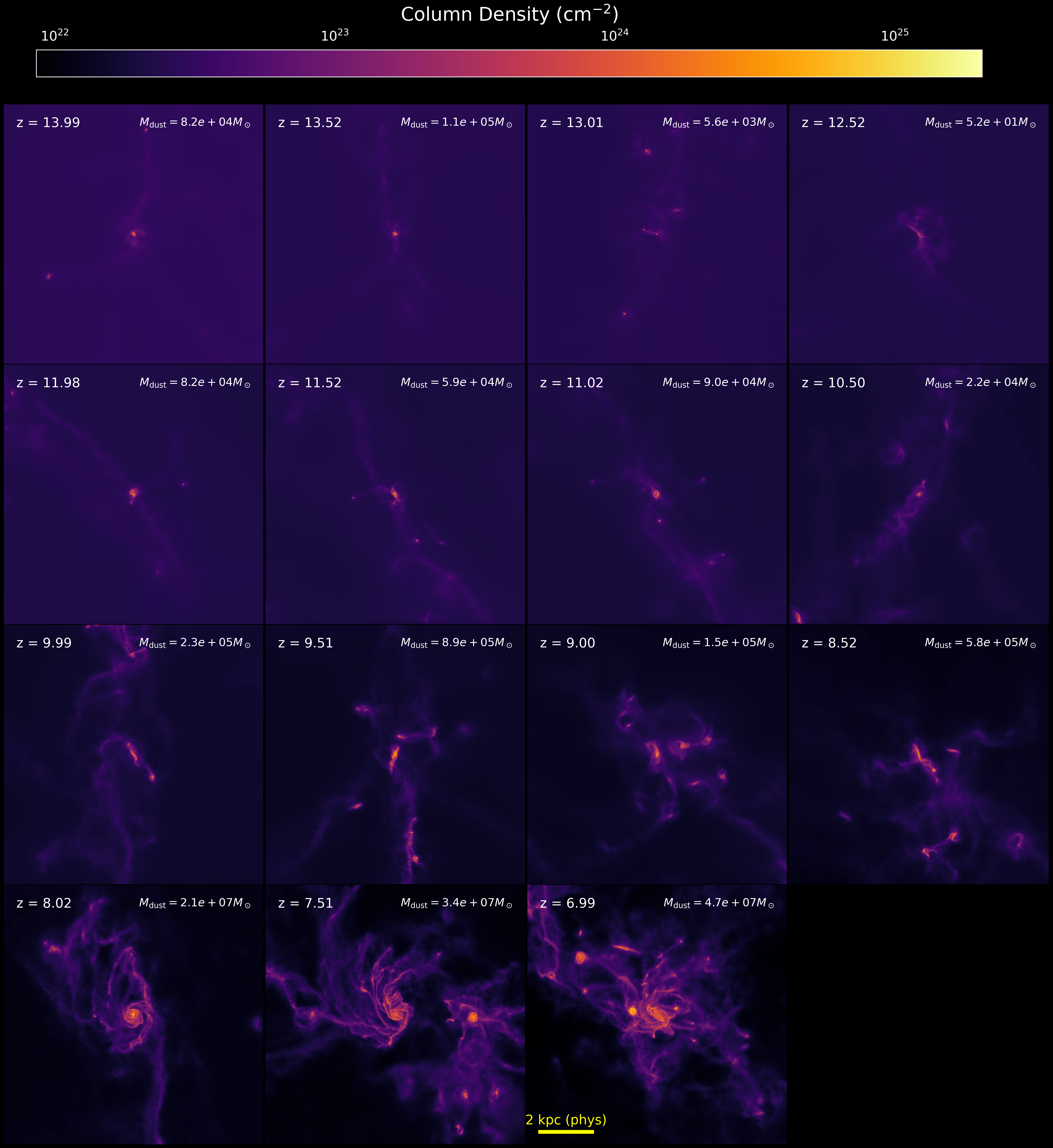}
    \begin{centering}
    \caption{{\bf Evolution of gas surface density of our fiducial example galaxy to show the morphological evolution of a
        massive galaxy in assembly at $z>7$}.  The subpanels are $10$
      kpc on a side, and the colorbar is in a fixed scale for all
      subpanels.  As expected, the galaxy grows hierarchically between
      $z=14\rightarrow7$, assembling via numerous minor mergers.  The
      mean density of the system increases dramatically as gas is
      funneled into the central galaxy; this rapid increase in density
      will prove important for the growth of dust in the first billion
      years.\label{figure:proj}}
    \end{centering}
  \end{figure*}

\begin{figure*}
  \centering
  \includegraphics[scale=0.4]{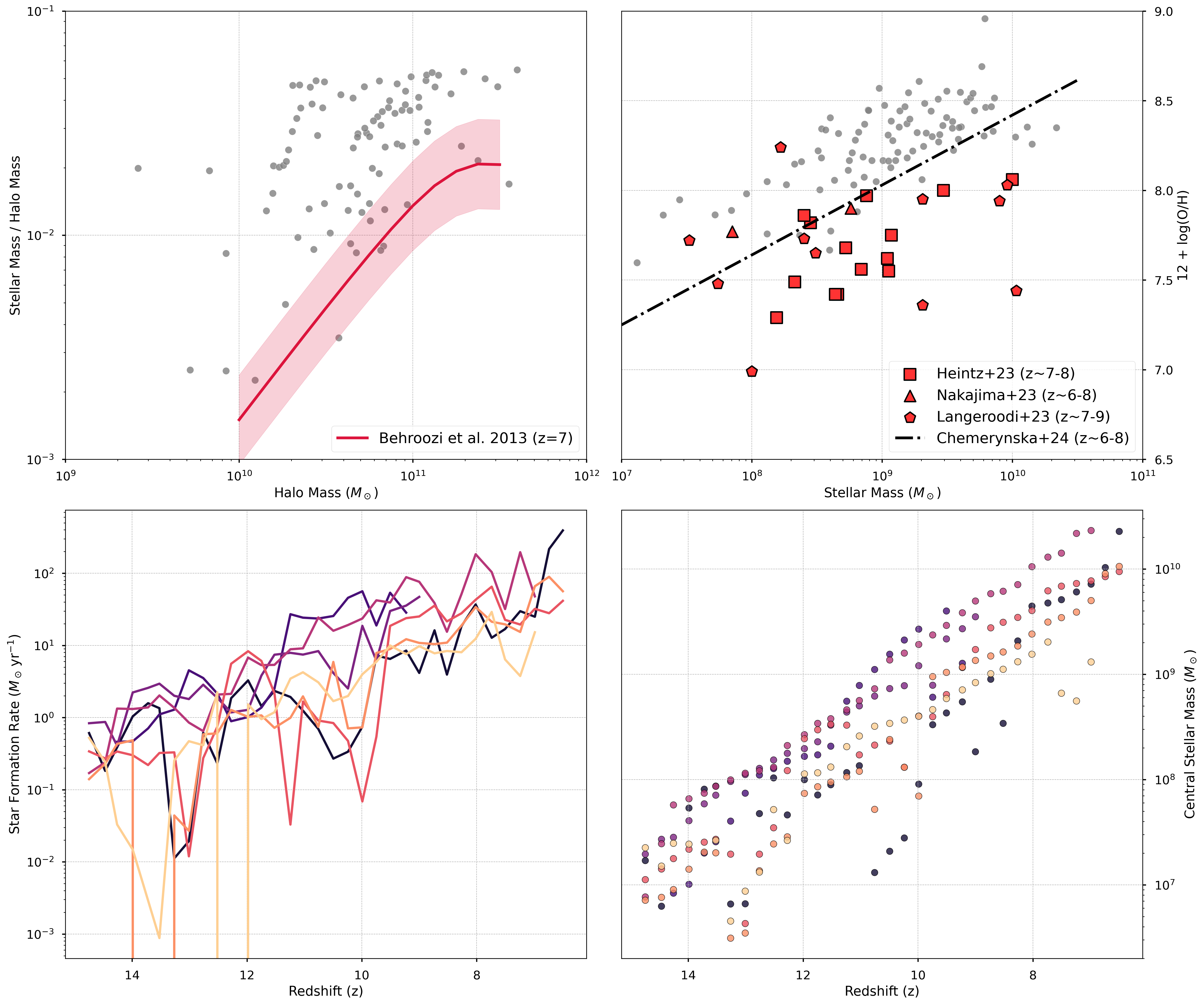}
  \caption{{\bf Physical properties of our model galaxies.}  Clockwise
    from top left: {\it Top Left:} The filled circles show the
    location of our model galaxies, selected between $z=7-8.5$ in
    stellar mass-halo mass space (each galaxy is represented by
    multiple points owing to multiple simulation snapshots between
    $z=7-8.5$).  The red shaded region shows constraints from the
      \citet{behroozi13a} semi analytic models and abundance matching
      techniques at $z=8$.  {\it Top Right:} Again, the filled
    circles show the location of our model galaxies, selected between
    $z=7-8.5$ compared to the observationally inferred $z \approx 6-8$
    mass-metallicity relationship (MZR).  The observational data for
    the MZR (red filled symbols, and dash-dot line) are from
    \citet{heintz23a,nakajima23a,langeroodi23a}, and the best fit line
    is from \citet{chemerynska24a}.  {\it Bottom Left and Right:}
    These panels show, respectively, the star formation history and
    stellar mass growth history of our model galaxies.  The color
    coding for the bottom two panels shows the individual
    galaxies. \label{figure:physical_properties}}
\end{figure*}

\section{Simulation Methodology}
\label{section:methods}
\subsection{Galaxy Physics and Zoom-In Technique}
We conduct a series of zoom-in
cosmological simulations of massive galaxies that are evolved down to
redshift $z_{\rm end} = 7$.  These simulations are conducted with the
{\sc smuggle} physics framework \citep{marinacci19a} within the {\sc arepo} codebase
\citep{springel10a,weinberger20a}.  Within {\sc smuggle}, we have
implemented a dust evolution scheme that is described in detail in
\citet{narayanan23a}, and we refer the reader to this paper for
the relevant equations and detailed description of the physics.
For the purposes of this work, we will briefly summarize at a high level the details of the
physics particularly relevant for this study.

We generate our initial conditions using {\sc music} \citep{hahn11a}.
We first run a low-resolution dark matter only simulation in a $(100
\ h^{-1} {\rm Mpc})^3$ box initialized at $z=99$ down to $z=0$.  We
identify dark matter halos using the {\sc caesar} galaxy analysis
package \citep{thompson14a}, and select a sample of massive
halos at $z=4$ with a diverse range of masses\footnote{This redshift was chosen for a balance of
computational efficiency (i.e., minimizing the number of particles
  that need to be split) with being reasonably well separated in time
from $z=7$ to minimize potential contamination by low-resolution
particles in our target halos. }.   In the Table, we list the masses of the halos at $z=4$ that were selected for resimulation at higher resolution.  These halos were picked arbitrarily to span a diverse range of masses.

\begin{table}
\label{table:table}
\centering
\caption{Halo masses at $z=4$ for target halos selected for high-resolution zoom-in simulations}
\label{table:halo_masses}
\begin{tabular}{cc}
\hline
Galaxy ID & $M_{\rm halo}$ ($M_\odot$) \\
\hline
1 & $1.02 \times 10^{13}$ \\
2 & $1.01 \times 10^{13}$ \\
4 & $7.73 \times 10^{12}$ \\
5 & $7.34 \times 10^{12}$ \\
250 & $1.29 \times 10^{12}$ \\
300 & $1.19 \times 10^{12}$ \\
350 & $1.10 \times 10^{12}$ \\
\hline
\end{tabular}
\end{table}

We first identify all particles within
$2.5 \times$ the radius of the maximum-distance dark matter particle
in the halo at $z=4$.
We then trace all particles within this region back to their initial $z=99$ locations, which we then use to generate a target high-resolution region mask. The
effective baryon mass resolution in the high-resolution region is initialized to $8.9 \times 10^4
M_\odot$.  Dust particles, when
present, have a factor $\sim 100$ lower mass on average.  We build a
sample of $6$ model galaxies to investigate via this technique, and
discuss their basic physical properties in
\S~\ref{section:physical_evolution}.  

Within these galaxies, gas cools via two-body collisional processes,
free-free emission, recombination, metal line cooling, and Compton
cooling off of the CMB \citep{katz96a}.  The line cooling rates are
computed as a function of density and temperature via {\sc cloudy}
photoionization models \citep{ferland13a,vogelsberger13a}.  At low
temperatures, alongside metal line cooling, fine structure and
molecular emission can be important: these rates are computed via fits
to {\sc cloudy} tables performed by \citet{hopkins18a}.  The
  abundance of metals available for cooling are self-consistently
  impacted by metal accretion onto dust grains (i.e. the metal
  abundances in gas cells is reduced/enhanced with
  dust grain growth/destruction). Gas self-shields at densities above
$n>10^{-3}$ cm$^{-3}$ \citep{rahmati13a} beginning at $z=5$, though self-shielding is not expected to play a significant role over the cosmic time interval considered in this paper.
Heating processes include cosmic rays \citep{guo08a}, and
photoelectric heating \citep{wolfire03a}.

Star formation occurs within gravitationally bound molecular gas.  We
set a density threshold for star formation of $n_{\rm thresh} \sim
150$ cm$^{-3}$.  We follow a volumetric \citet{kennicutt98a} star
formation relationship, with an efficiency per free fall time of
$\epsilon_{\rm ff} \approx 0.01$ \citep{krumholz07b}.  The molecular
fraction of neutral gas is computed via the \citet{krumholz08a}
prescriptions balancing Lyman-Werner photodissociation rates against
molecular formation rates.  These stars return energy and momentum
back into the ISM via a range of channels, including stellar mass loss
and energy injection by Type Ia and II SNe
\citep{vogelsberger13a,torrey14a,marinacci19a,zhang24a},
photoionization by massive stars, radiation pressure, and OB and AGB
stellar winds \citep[see ][for details]{marinacci19a}.  The radiation pressure does not directly couple to the dust particles, though they are impacted by feedback events via gas-dust drag.  Models directly coupling the radiation to the dust will be presented in future work.  Metals that
are injected into the ISM are advected as passive tracers with the
fluid flow, which allows for a self-consistent coevolution of galaxies
and their metal content.  We discuss the metallicities of our model
galaxies in the context of observations in
\S~\ref{section:physical_evolution}.  It is these free metals in the
ISM that constitute the reservoir of material available for accretion
onto dust grains.

\subsection{Dust Physics}
We follow the live dust formalism developed by
\citet{mckinnon16a,mckinnon17a,mckinnon18a}, including updates
for multi-size and multi-composition dust grains by \citet{li21a} and
\citet{narayanan23a}.  Dust forms in the ejecta of evolved stars.
Specifically, a fraction of metals returned from SNe and AGB stars is
assumed to condense into dust, with yields derived from
\citet{schneider14a} for AGB stars, and \citet{nozawa10a} for SNe, and this dust 
is initialized with a grain size distribution.  Hydrodynamic models
\citep{winters97a,yasuda12a} suggest that the size
distribution from AGB stars is reasonably well represented by a
lognormal distribution.  While there are subtle differences in the location of the
peak, both models agree that large grains dominate, with the peak $a_0
\sim 0.1-0.2 \mu$m.  Similarly, \citet{nozawa07a} compute the size
distribution for dust grains formed in Type II SNe, and also suggest
similarly large grains formed upon formation (owing to destruction of
small grains by the reverse shock).  Following \citet{asano13a}, we
therefore assume lognormal size distributions for the initialized dust populations formed in SNe and AGB stars:
\begin{equation}
\frac{\partial n}{\partial {\rm log}a} = \frac{C}{a^p}{\rm exp}\left(\frac{{\rm ln}^2\left(a/a_0\right)}{2\sigma^2}\right)
\end{equation}
where $C$ is a normalization constant, and $a_0 = 0.1 \mu$m,
$\left(p,\sigma\right) = \left(4,0.47\right)$ for AGB produced dust,
and $\left(p,\sigma\right) = \left(0,0.6\right)$ for SNe.  $\sigma$ determines the width of the distribution, and $p$ the power-law behavior at the extremes.  As we will
discuss later in this paper, the grain size distribution can play an
important role in dust growth in the ISM: this said, the dust
particles lose their memory of their initialized size distribution
rapidly, and our main results are not sensitive to this assumption.  We show this explicitly in the Appendix.

Following the methodology of \citet{mckinnon18a}, we model the
  interstellar dust population using a distinct set of simulation
  particles that allow it to move independently from the gas, except
  for explicitly implemented gas-dust interaction terms. In this
  framework, dust is represented by discrete simulation particles that
  are numerically decoupled from the gas mesh. Unlike models that
  treat dust as a passive scalar fixed to the gas flow
  \citep[e.g.][]{mckinnon16a,mckinnon17a,li21a,esmerian22a,parente22a,choban22a},
  our live dust particles are subject to independent gravitational
  forces and a semi-implicit aerodynamic drag force. This allows for
  resolved dust--gas relative motion, particularly in low-density
  regions where the stopping timescale $t_s$ is long, while
  maintaining accurate coupling in the high-drag regime of the dense
  ISM.

Once formed, dust can grow via the accretion of free metals.
Formally, the growth timescale is inversely proportional to the
density of available metals in the nearby ISM, as well as the square
root of the temperature.  There is additionally a dependence on the
size of the dust grain: smaller dust grains have a larger surface area
per volume than large grains.  Formally, we model the rate of dust accretion timescale as:
\begin{equation}
\left(\frac{da}{dt}\right)_{\rm grow} = \frac{a}{\tau_{\rm accr}}
\end{equation}
where $a$ is the size of the dust grain, and $\tau_{\rm accr}$ is the accretion timescale, given by:
\begin{equation}
  \label{equation:growth}
\tau_{\rm accr} = \tau_{\rm ref}\left(\frac{a}{a_{\rm ref}}\right)\left(\frac{\rho_{\rm ref}}{\rho_{\rm g}}\right)\left(\frac{T_{\rm ref}}{T_{\rm g}}\right)^{1/2}\left(\frac{Z_{\rm ref}}{Z_{\rm g}}\right)\left(\frac{S_{\rm ref}}{S}\right).
\end{equation}
Here, $\rho_{\rm g}$, $T_{\rm g}$, $Z_{\rm g}$ and $S$ are the gas
density, temperature, metallicity, and grain-metal sticking
coefficient, respectively. $\tau_{\rm ref}$, $a_{\rm ref}$, $\rho_{\rm
  ref}$, $T_{\rm ref}$, $Z_{\rm ref}$ and $S_{\rm ref}$ are reference
values. Following \citet{narayanan23a}, we adopt the reference values:
$\tau_{\rm ref} = 0.224$ Gyr for silicates and $\tau_{\rm ref} =
0.175$ Gyr for carbonaceous grains, $\rho_{\rm ref} = 100 m_{\rm H}$
cm$^{-3}$, $a_{\rm ref} = 0.1 \mu$m, $S_{\rm ref}=0.3$, $T_{\rm ref} =
20$ K, and $Z_{\rm ref} = Z_\odot = 0.0134$.  Following
\citet{zhukovska16a}, we additionally adopt a temperature dependent
sticking coefficient which drops at higher temperatures.  This
  model assumes that grain growth is efficient in the cold ISM ($T \le
  300$\,K) but is physically suppressed at higher temperatures where
  thermal energies inhibit the sticking of gas-phase species to grain
  surfaces (this is a smoothed approximation to the
  \citet{zhukovska16a} step function model). This ensures that the
  majority of dust growth in our simulations occurs within the cold
  neutral medium and molecular clouds.

Dust can be destroyed via thermal sputtering (i.e., the erosion of
dust grains) in the ISM, and via SNe shocks in star-forming regions.
The sputtering timescale is adopted from \citet{tsai95a}, and is
proportional to the grain size and inversely proportional to the local
gas temperature and density:
\begin{equation}
  \label{equation:sputtering}
  \left(\frac{da}{dt}\right) = -\frac{a}{\tau_{\rm sp}}
\end{equation}
where
\begin{equation}
    \tau_{\rm sp} = \left(0.17 {\rm Gyr}\right)\left(\frac{a}{a_{\rm ref}}\right)\left(\frac{10^{-27} {\rm g \ cm^{-3}}}{\rho_{\rm g}}\right)\left[\left(\frac{T_0}{T}\right)^\omega+1\right].
\end{equation}
Here, $\omega=2.5$ controls the low-temperature scaling of the
sputtering rate, and $T_0 = 2 \times 10^6$ K.  The SNe destruction
rate is modeled following the dust destruction model near blast waves
by \citet{nozawa06a} and \citet{asano13a}.  The explicit equations for
these size transformation processes are given in \citet{narayanan23a}.

Finally, dust grains can change size owing to interstellar collisions.
Here, there are two primary effects: dust shattering, which occurs in
high-speed collisions between dust grains \citep{jones96a}, and
coagulation (i.e., sticking together) which can occur in lower speed
collisions.  The former process transforms large grains into smaller
grains, while the vice versa is true in coagulation.  We model the
transformation of grain sizes in collisions following
\citet{mckinnon18a} and \citet{li21a}.

Quantitatively, we model the transformation of grain sizes by the mass evolution of grain-size bin $k$:
\begin{equation}
  \label{equation:shattering}
    \begin{aligned}
      \frac{{\rm d}M_k}{{\rm d}t} = -\frac{\pi \rho_{\rm d}}{M_{\rm d}} \Bigg( \sum_{k=0}^{N-1} v_{\rm rel}(a_i,a_k) m_i I^{\
i,k} - \\
      \frac{1}{2} \sum_{k=0}^{N-1} \sum_{j=0}^{N-1} v_{\rm rel} (a_k,a_j)  m^{k,j}_{\rm col}(i) I^{k,j}\Bigg)
  \end{aligned}
    \end{equation}
as long as the relative velocity between grains is greater than a
threshold velocity $v_{\rm rel} > v_{\rm thresh}$.  Here, the grain
sizes are denoted with $a$, $m_i$ is the mass of the grain in bin $i$,
and $m^{k,j}_{\rm col}\left(i\right)$ is the resulting mass entering
bin $i$ due to the collision between the grains in bins $k$ and $j$.
We employ threshold velocities of $v_{\rm thresh} = 2.7$ km s$^{-1}$
for silicates, and $1.2 $ km s$^{-1}$ for carbonaceous grains based on
the modeling of \citet{jones96a}.  The transfer of mass between size
bins is analogous for coagulation, and is implemented if $v<v_{\rm
  thresh}$.  We follow \citet{hirashita09a} in employing a threshold
velocity that is dependent on the grain sizes as well as the material
properties of the species, following their Equation~8.

The relative velocities of grain-grain collisions are computed
  {\it within} dust super particles, and are governed by a
  subresolution model for unresolved turbulence.  We assume that grain
  motions within individual dust super-particles are driven by a
  turbulent cascade where the energy injection scale, $L_{I}$, is
  taken to be the local Jeans length. The relative velocity between
  grains is calculated based on their respective aerodynamic coupling
  to this unresolved turbulent gas flow. Specifically, the relative
  velocity of a grain of size $a$ with respect to the gas is a
  function of its stopping time, $t_s$, which depends on the local gas
  density, grain size, and sound speed.

We note that
while there is of course some freedom in choosing the critical
velocities for either coagulation or shattering, these default values
have resulted in dust grain size distributions comparable to
\citet{mathis77a} ``MRN'' distributions for Milky Way-like zoom in
galaxies \citep{li21a}, as well as reasonable model PAH spectra
\citet{narayanan23a}. As a result, we adopt these values without any
tuning, though note that in a future work, we anticipate presenting
the impact of a parameter exploration on the modeled dust properties
of simulated galaxies.


\begin{figure}
  \includegraphics[scale=0.5]{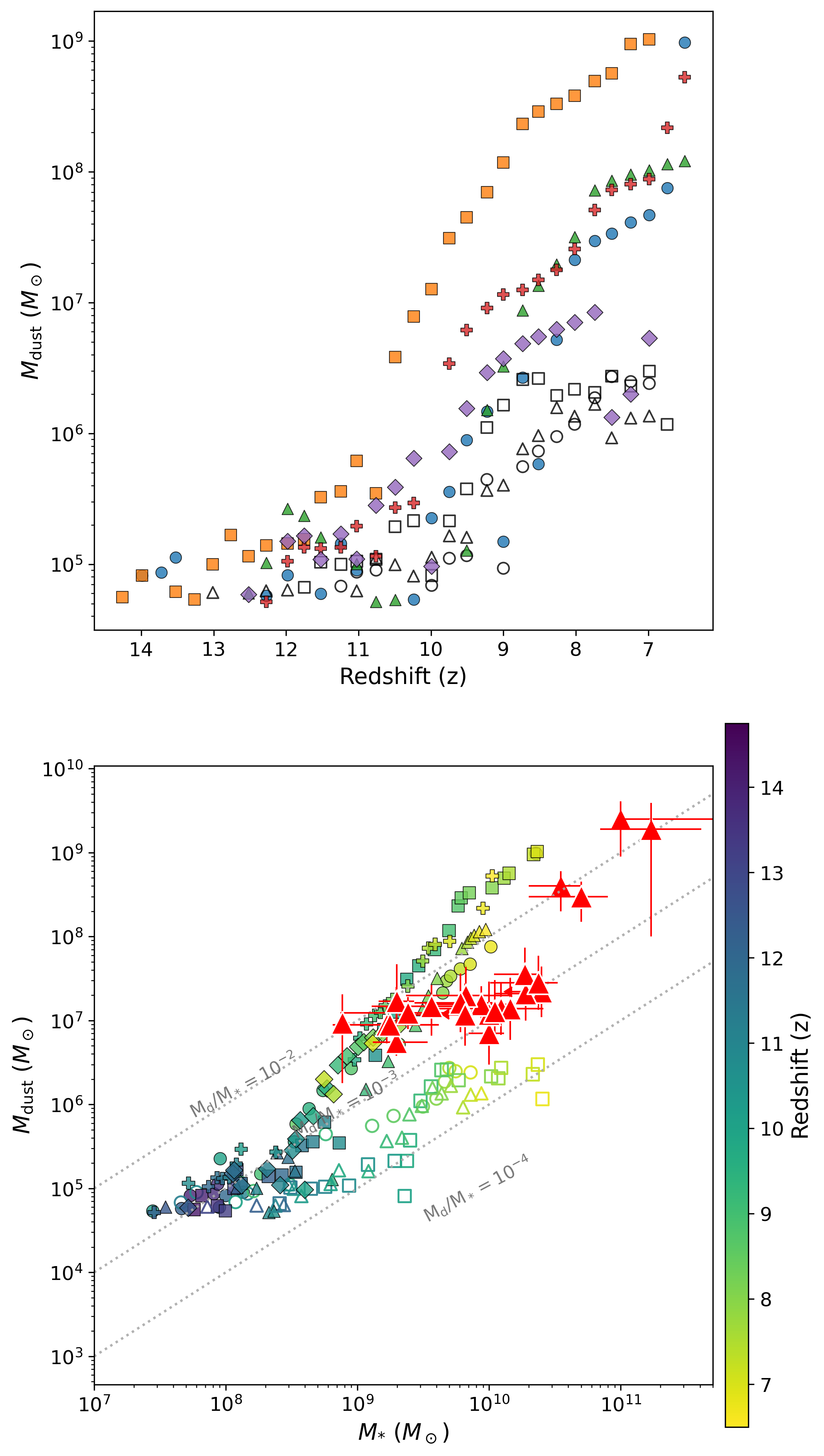}
  \caption{{\bf Dust growth takes over from production only mechanisms
      by $z\sim 10-11$, and is necessary in order to match the
      observed galaxy dust masses at $z \sim 6-7$.} {\it Top:} $M_{\rm
      dust}$ vs $z$ for all of our model galaxies.  {\it Bottom:}
    $M_{\rm dust}$ vs $M_*$ for all model galaxies, color-coded by
    redshift (with observations at $z=6-8$ shown with red triangles).
    In both panels, the filled symbols are our fiducial model, with
    different shapes corresponding to different model galaxies, while
    the open symbols represent a numerical experiment in which we turn
    off dust growth.  Production only models (open symbols) are able
    to achieve $M_{\rm dust}/M_*$ ratios of $\sim 10^{-4}$, though to
    reach the dust masses inferred by observations, rapid dust growth
    is necessary.  The observational data (red triangles) come from
    \citet{cooray14a,watson15a,laporte17a,marrone18a,hashimoto19a,fudamoto21a,ferrara22a}
    and \citet{topping22a}. \label{figure:dust_grow}}
  \end{figure}

\section{The Physical  Properties of Early Universe Galaxies}
\label{section:physical_evolution}

We now turn to examining the basic physical properties of our model
galaxies.  To help place the galaxy formation model in context, in
Figure~\ref{figure:proj}, we show the evolution of the gas surface
density of a model galaxy (Galaxy 1 in Table \ref{table:halo_masses}) between $z=14\rightarrow7$.  The individual
subpanels are $10$ kpc on a side.  The central galaxy assembles rapidly, forming a relatively compact, dense galaxy by $z \approx 7$.  Throughout
this paper, in cases where we show an individual example galaxy (as
opposed to the aggregate properties of all of our $6$ model galaxies),
we show the same galaxy depicted in Figure~\ref{figure:proj}.

 In Figure~\ref{figure:physical_properties}, we show the location of
 our $6$ model galaxies in stellar mass-halo mass space, the
 mass-metallicity relationship (MZR), and their redshift evolution of
 star formation rate and stellar mass.  This Figure is intended to
 demonstrate correspondence between our modeled galaxy physical properties, and those inferred from observations of high-$z$ galaxies.   We discuss the individual subpanels of
 Figure~\ref{figure:physical_properties} in turn.

The solid line in the stellar mass halo mass relation are inferred
$z=8$ constraints by \citet{behroozi13a}, while the filled circles are
our model galaxies within a relatively narrow redshift range
$z=7-8.5$.  In the top right panel, we show observational constraints
on the mass metallicity relationship \citep[filled red
  points;][]{heintz23a,nakajima23a,langeroodi23a}, and the best fit
relation from \citet{chemerynska24a} as a dash-dot line.  Both
relationships show reasonable correspondence between our model
galaxies and observational inferences, though we note that our model
galaxies have modestly higher stellar masses per halo mass, and higher
metallicities per stellar mass than observational constraints at the
same epoch.  Given the relative uncertainty in estimating stellar
masses at high-$z$
\citep[e.g.][]{topping22a,lower23a,whitler23a,narayanan24a}, as well
as metallicities from strong line measurements
\citep{garg23a,chakraborty25a,vijayan25a,sanders25a}, we are
encouraged by the relative correspondence between our simulations and
observational constraints\footnote{We note that if our simulated
stellar mass-halo mass or mass-metallicity relations in reality exceed
those of real galaxies at $z \approx 7-8$, then this will have the
consequence in our paper of the transition to a growth dominated phase
of dust as occuring later in reality than our models suggest.  This
said, because the transition in our models from production dominated
to growth dominated is a consequence of the local physical conditions
in the ISM, the underlying physics of dust growth and the origin of
Blue Monsters in this study remains unchanged, even if the absolute
timing of this maturation phase is earlier in our models than reality.}.  The high dispersion in the
simulated MZR is a natural consequence of bursty star formation
histories \citep{ma16a,bassini24a,garcia24a,menon25a,mcclymont25a}.
Generally, within the scope of observational uncertainties in galaxy
property inference at these epochs, we consider our model galaxies to
be viable with respect to observational constraints.

We now turn to the redshift evolution of the star formation rates and
stellar mass.  The star formation histories of these model galaxies
are bursty.  Bursty star formation, especially at relatively low
galaxy masses, is an expected outcome for simulations that include
explicit feedback models, such as {\sc fire} and {\sc smuggle}
\citep{hopkins14a,sparre17a,marinacci19a,gurvich23a,shen23a,sun23a,mcclymont25a,narayanan24a,narayanan25a,shen25a}.
While the SFHs of high-$z$ galaxies are still uncertain, at least some
observational inferences suggest that such a SFH may be plausible
\citep{dressler23a,endsley23a,shen23a,asada24a,ciesla24a,clarke24a}.
Similarly, while there are few (and generally, uncertain) mass
constraints on early Universe galaxies, we note that the stellar
masses of our model galaxies at $z \approx 7$ are in
reasonable correspondence with typical ALMA detections 
\citep[e.g.][]{topping22a}.

Taken together, the reasonable correspondence between our model and
the stellar mass-halo mass relation, the mass-metallicity
relationship, the model star formation histories, and model stellar
masses with observational inferences is a statement that the
combination of stellar feedback models, star formation, and metal
consumption by dust are appropriate for these simulations.  While there
are of course likely other parameter combinations in our subresolution
modeling that would {\it also} result in acceptable matches to these
observed relationships, the correspondence seen in the top row of
Figure~\ref{figure:physical_properties} gives us confidence that we
may treat these galaxy simulations as test beds for numerical
experiments, studying the rise of dust in the early Universe.


\section{The Growth of Dust in High Mass Halos in the Early Universe}
\label{section:dust_growth}

We now turn to the first major question that this paper aims to
address: how does dust grow rapidly in the early Universe?  A key
point of this paper will be to demonstrate the importance of dust
growth in the ISM for achieving relatively large dust-to-stellar
ratios (i.e. $\sim 10^{-2}$).  At the same time, dust yields for SNe
and AGB (i.e., ``production-only'' mechanisms for dust content in
galaxies) are fairly uncertain \citep[see ][for a recent
  review]{schneider24a}.  To anchor our results going forward, we have
computed the maximum expected dust to stellar mass ratio from
  stellar production only, finding $M_{\rm dust}/M_* \approx 3 \times
10^{-3}$, with details of the calculation presented in the Appendix.
We note that this is an upper limit for the expected dust-to-stellar
mass ratio expected from production, and does not take into account
potential destruction near sites of star formation.


Building from this estimate, in Figure~\ref{figure:dust_grow}, we plot the redshift
  evolution of the dust mass of all of our model galaxies (top panel),
  and the location of the model galaxies on the dust mass-stellar mass
  plane (bottom-panel).  Individual points are color-coded by
  redshift.  The value in the latter plot is that it can more easily
be directly compared to observational constraints.

For each plot, we show our fiducial runs with filled symbols.  We
additionally show the results of a numerical experiment that we have
run in which we have turned off the physics of dust growth, and denote these galaxies via open
symbols.  We first examine our fiducial model, which includes the
physics of dust growth.  Figure~\ref{figure:dust_grow} demonstrates
$3-4$ orders of magnitude increase in dust mass between
$z=14\rightarrow7$ in our fiducial model galaxies.  While there can be
minor fluctuations from monotonic growth (owing to feedback events
disrupting relatively low-mass halos in the early Universe), there is
generally a rapid increase of dust masses in the ISM of massive
galaxies in the early Universe.  The simulated galaxies display
both dust masses comparable to ALMA detections at $z\sim6-7$, as well
as dust to stellar mass ratios of $\sim 10^{-3}-10^{-2}$.

\begin{figure}
  \includegraphics[scale=0.375]{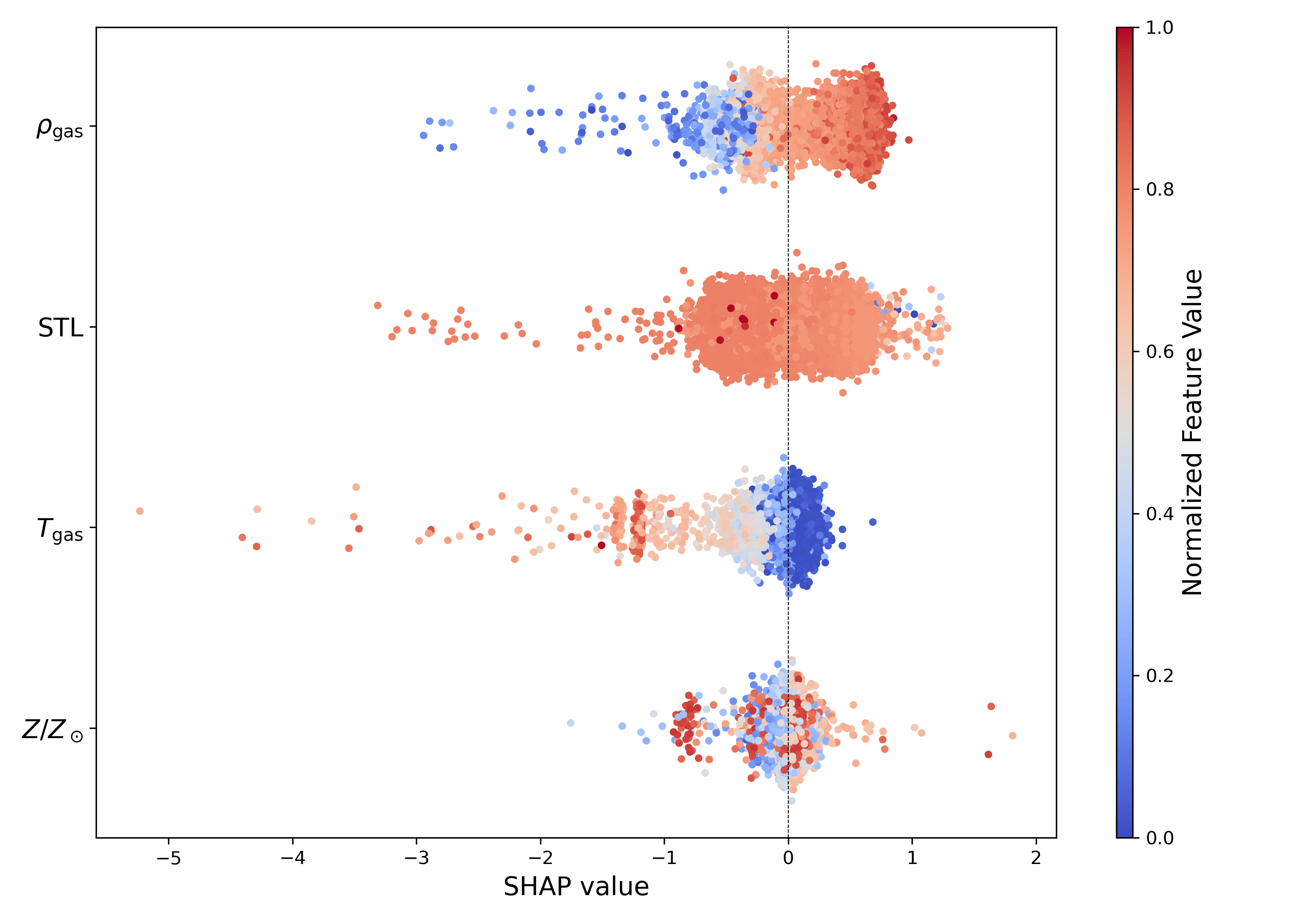}
\caption{{\bf Relative
    importance of individual physical components to the dust growth in
    early Universe galaxies.}  The rows show the results of a
  \citet{shapley53a} analysis (with the details presented in
  \S~\ref{section:physical_drivers}) of the individual dust particles
  in our fiducial example galaxy over all redshifts, and are ordered by the most
  important ingredients to least important for dust growth.  The
  increase in dust density is the dominant feature in driving dust
  growth (Figure~\ref{figure:proj}), while the increase in the number
  of small grains is the second most important
  (Figure~\ref{figure:gsd}).  Positive \citet{shapley53a} (SHAP) values denote a strong correlation between a given physical property and the label (dust mass), while negative SHAP values denote an anticorrelation.  Red swarms correspond to large values of a given physical quantity (whose normalized ranges are shown in the colorbar), whereas blue swarms correspond to small values.   The vertical dispersion for a given physical quantity is simply imposed jitter to clarify the location of the points.  See \S~\ref{section:physical_drivers} for a detailed interpretation, including understanding covariances.  \label{figure:shapley}}
\end{figure}

\begin{figure}
  \includegraphics[scale=0.25]{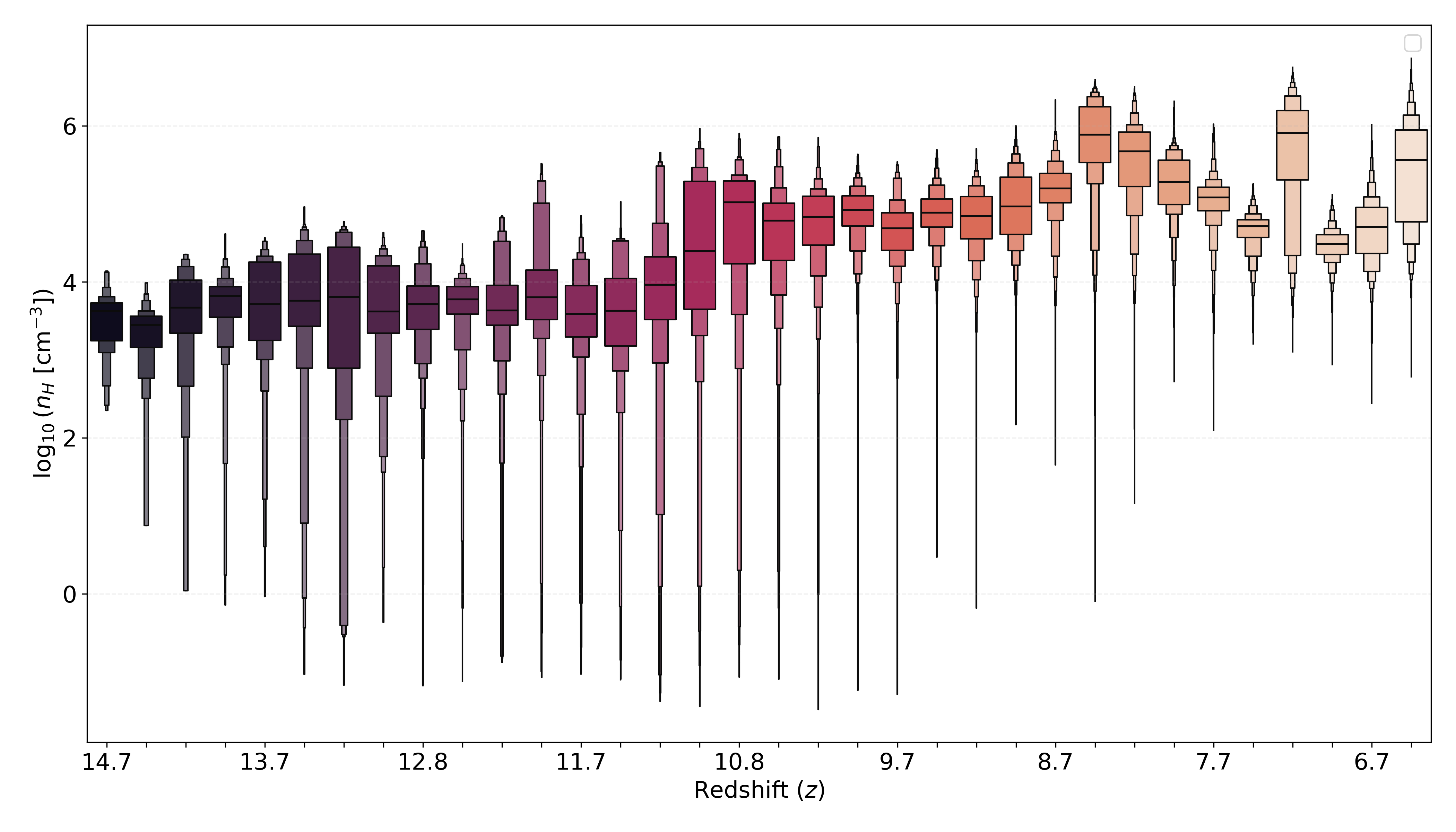}
  \caption{{\bf Distribution of gas densities in the vicinity of dust particles as a function of redshift.}  The boxen plots represent the mass weighted gas density near dust particles in our model galaxies.  For each redshift bin, the widest central box denotes the interquartile range (25$^{\rm th}$-75$^{\rm th}$ percentile), while subsequently narrower boxes denote deeper quartile ranges (12.5$^{\rm th}$, 6.25$^{\rm th}$, etc.). \label{figure:dust_dens_z}}
\end{figure}

\begin{figure*}
  \includegraphics[scale=0.4]{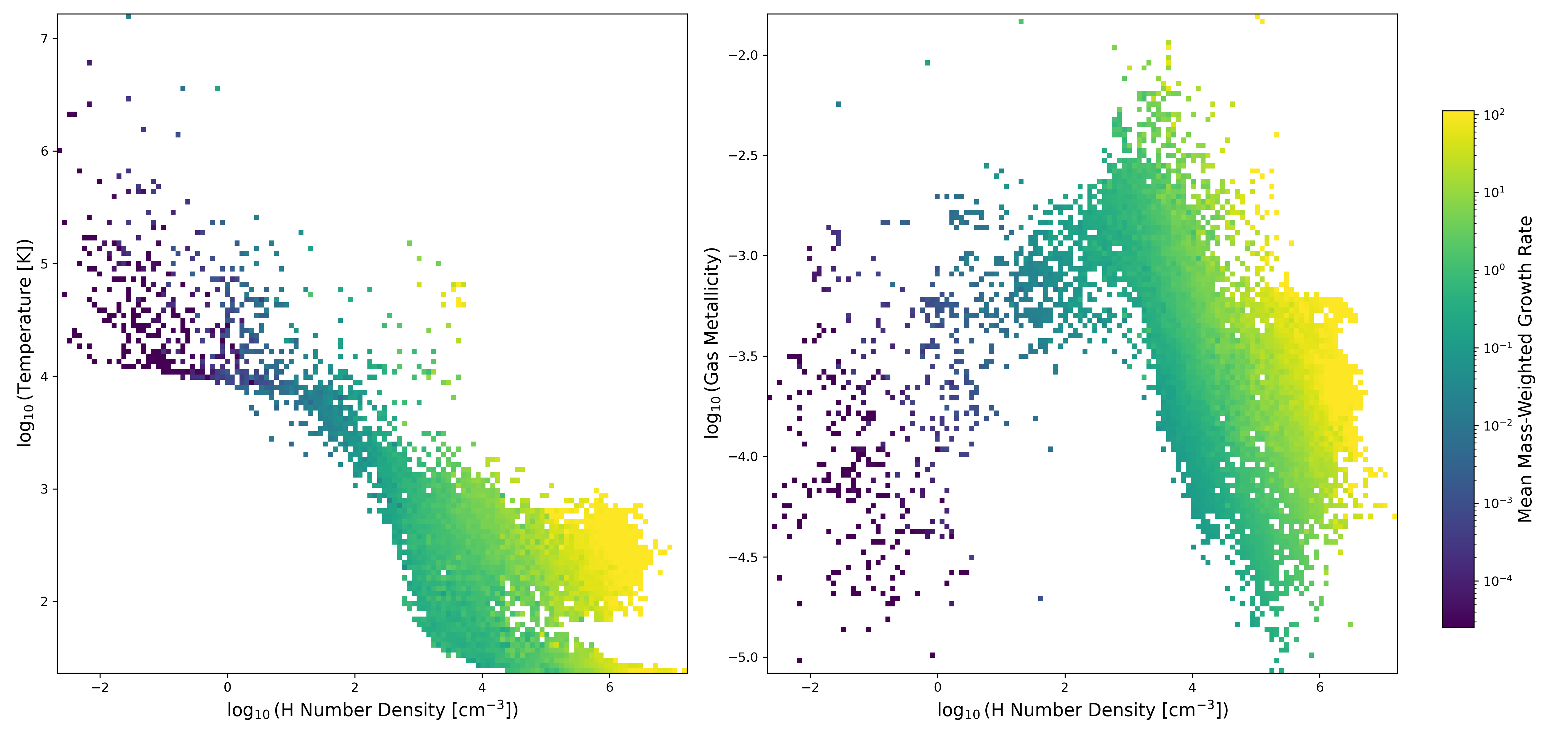}
  \caption{{\bf 2D Histograms of physical properties of dust particles in our fiducial example galaxy showing that the main dust growth is concentrated near high density, moderate temperature gas.}  The main purpose of this Figure is to aid in the interpretation of Figure~\ref{figure:shapley}, and the associated discussion in \S~\ref{section:physical_drivers}. \label{figure:2dhist}}
\end{figure*}

The rapid dust mass increase owes primarily to the growth of dust
grains via metal accretion.  This is clear when examining the results
of our production-only numerical experiment (open symbols) in
Figure~\ref{figure:dust_grow}.  At early times ($z \ga 10-11$), dust
production from SNe dominates the dust budget of massive galaxies.
This can be seen by the general correspondence of dust masses of our
model galaxies when the physics of growth is included, and neglected
in the top and bottom of Figure~\ref{figure:dust_grow}.  Dust
production alone by SNe is enough to result in a dust to stellar
mass ratio $\sim 10^{-4}$, in good agreement with the
analytic derivations for the expected dust content from production alone by \citet{ferrara25a}.  This said, at $z\la10-11$, for the
galaxy masses modeled here, metal accretion is necessary to drive the
rapid rise in dust content.  Once dust growth begins to dominate,
there is a rapid departure between the fiducial model and no
growth model's dust content in both panels of
Figure~\ref{figure:dust_grow}.  This results in the dust masses
growing sufficiently large to reach dust-to-stellar mass ratios
$M_{\rm d}/M_* \ga 10^{-2}$, comparable to inferred values from
observations from $z \sim 6$ ALMA detections \citep{cooray14a,watson15a,laporte17a,marrone18a,hashimoto19a,fudamoto21a,ferrara22a,topping22a}.


\subsection{The Key Physical Drivers of Dust Growth in the Early Universe}
\label{section:physical_drivers}

Having demonstrated the importance of dust growth in achieving the
rapid rise in dust masses in galaxies in the first billion years, the
next key question is: what physical process drives the growth of this
dust?  Recalling Equation~\ref{equation:growth}, there are $4$ major
ingredients that contribute to the growth timescales: dust grain sizes, local gas density,
metallicity, and temperature.

To isolate the driving physical cause of early dust growth, we have
trained a machine learning algorithm ({\sc xgboost}), where the labels
are the dust masses of individual dust particles, and the features are
the key physical components of the growth equation: the dust size
distribution, and temperature, density, and metallicity of neighboring
gas particles.  The model is trained on the dust particles of every galaxy snapshot between $z=7-14$.  Once trained\footnote{In the Appendix, we demonstrate the reliability of this machine learning model.}, we perform a \citet{shapley53a} feature
importance analysis.  The premise behind \citet{shapley53a} indices
(hereafter, Shapley values or SHAP values) is based on game shows
where the teams collaborate with one another against a common challenge for a prize, asking
the basic question: if different members of the team contribute
differently, how should the prize be divided up?  The analogy to our
model is straight forward: the physical features act as the `players,'
and the `prize' is the prediction for an individual particle's dust
mass. For any given dust particle, the Shapley value quantifies how much
each physical feature contributed to pushing the model's prediction
away from the average prediction for the entire dataset. A key
advantage of this method is its ability to fairly distribute these
contributions even when the features are highly
interdependent\footnote{See, e.g \citet{gilda21a} for an example
use-case of Shapley values in galaxy SED fitting.}, as is the case in
our simulations where gas density, temperature, and metallicity are
all correlated. By aggregating the Shapley values from all particles
in our sample, we can build a global picture of feature importance,
revealing not only which physical quantities are most predictive but
also the nature and direction of their impact on dust growth.

\begin{figure}
  \includegraphics[scale=0.25]{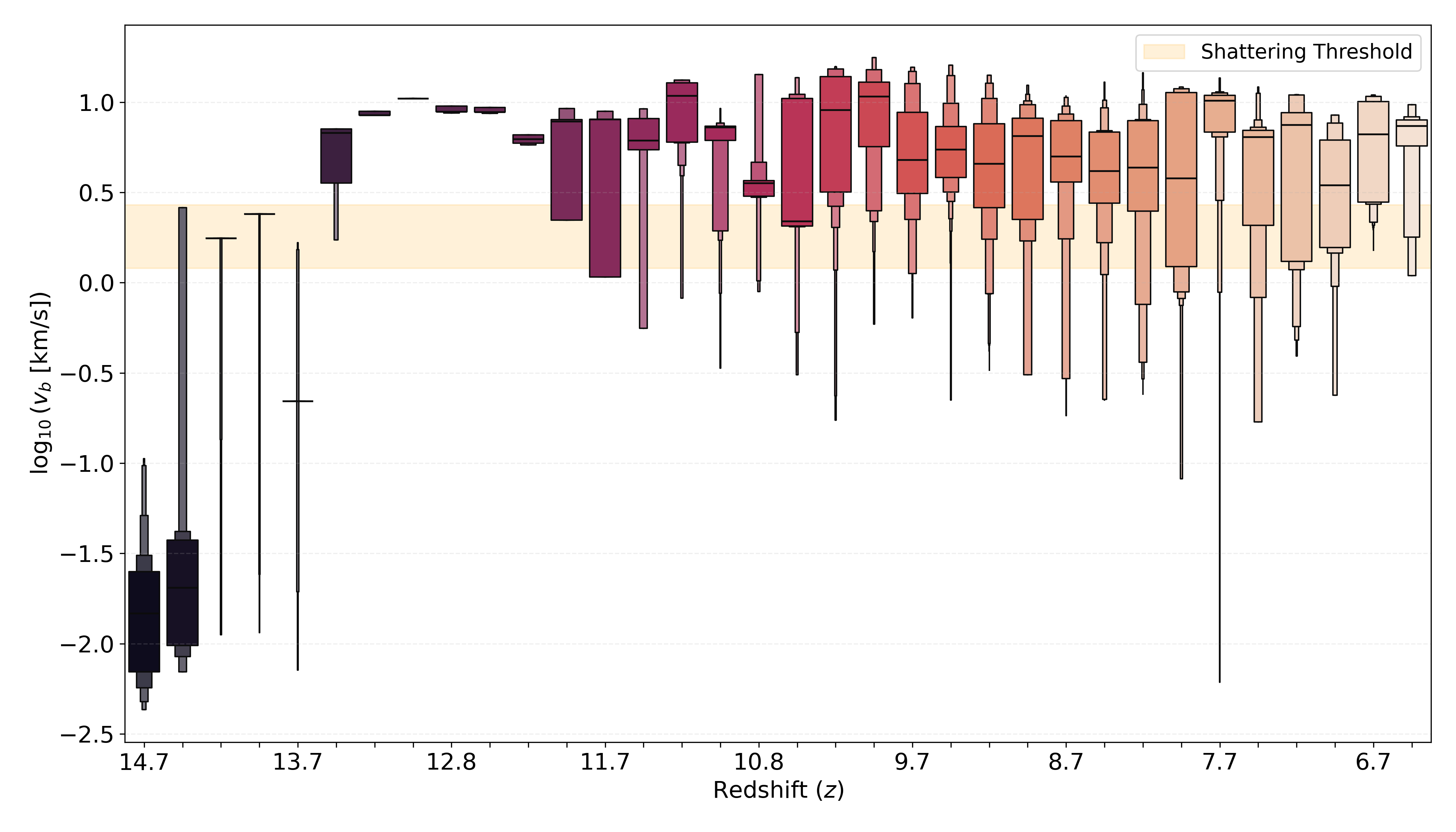}
  \caption{{\bf Distribution of collision velocity distribution as a
      function of redshift.}  The boxen plots represent the grain number
    weighted velocity distribution, while the shaded orange region indicates the shattering threshold velocities implemented in our model.  For each redshift bin, the widest central box denotes the interquartile range (25$^{\rm th}$-75$^{\rm th}$ percentile), while subsequently narrower boxes denote deeper quartile ranges (12.5$^{\rm th}$, 6.25$^{\rm th}$, etc.). \label{figure:vdisp_z}}
\end{figure}

\begin{figure*}
  \centering
  \includegraphics[scale=0.6]{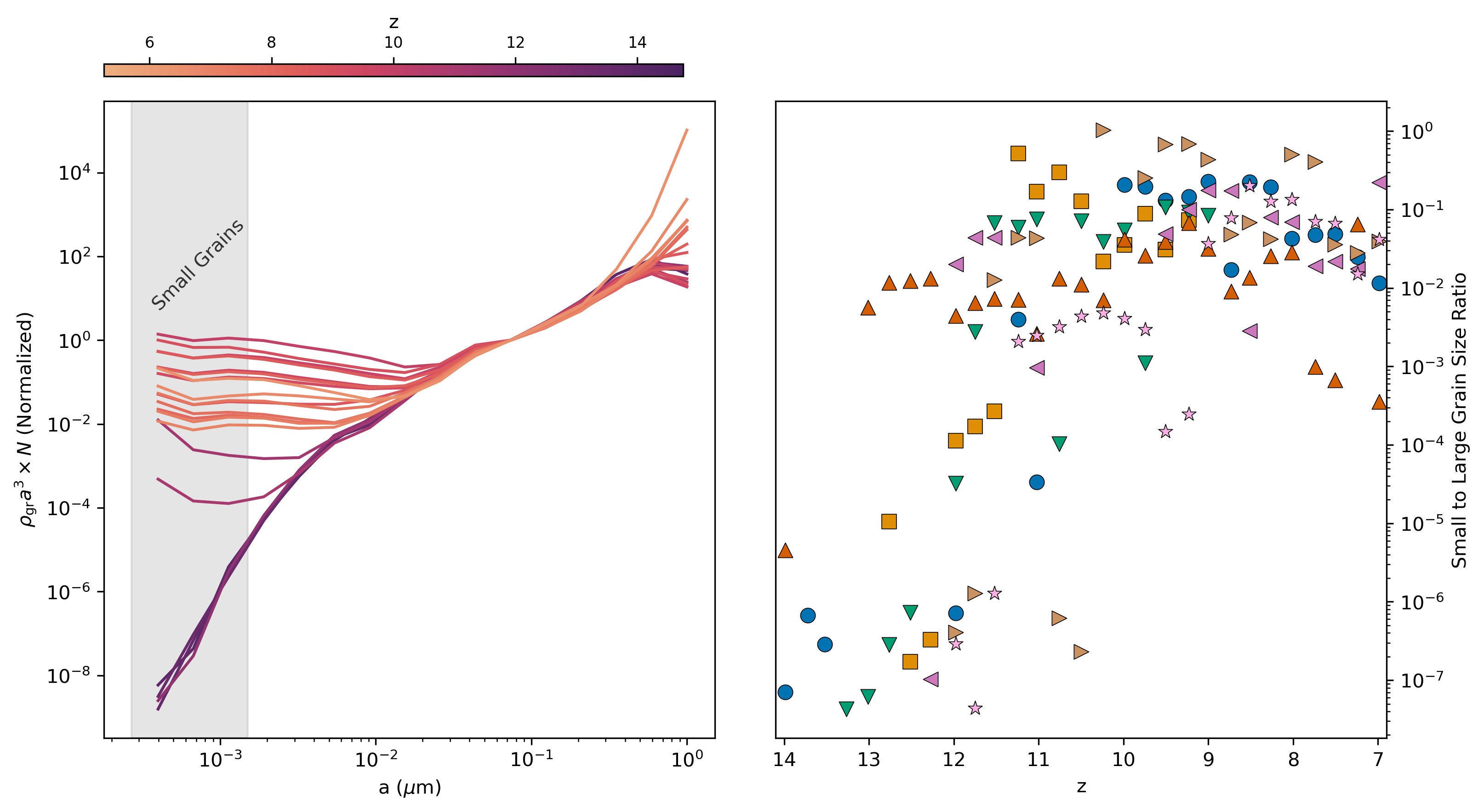}
  \caption{{\bf The fraction of small dust grains grows dramatically
      as redshift decreases in the first galaxies.} In the left, we
    show the redshift evolution of the dust grain size distribution
    for our fiducial example galaxy, showing the shifting of power
    from large to small grains.  This owes to an increase in the grain-grain shattering rates as the interstellar turbulent velocities increase as the halo assembles and star formation rate increases for the galaxy.  In the right panel we show the ratio of total small grains to large grains as a function of redshift, with the demarcation between small and large grains shown in the left panel.  Generally, the small to large ratio increases as redshift decreases for all model galaxies. \label{figure:gsd}}
\end{figure*}

In Figure~\ref{figure:shapley}, we present this feature importance
decomposition for our fiducial example galaxy between redshifts $z=7-14$. The features are
ranked on the ordinate by their global importance, determined by the
mean absolute SHAP value across all particles. On the abscissa, the
SHAP value represents the contribution of that feature to the model's
prediction for a single particle. A positive SHAP value indicates a
contribution that increases the predicted dust mass, while a negative
value indicates a contribution that decreases it. The magnitude of the
SHAP value corresponds to the strength of this contribution. Finally,
the color of each point corresponds to the value of the feature
itself, with red denoting high values (e.g., high gas density) and 
blue denoting low values.  Because the individual physical properties all have different ranges of values, in order to collapse this information into a single colorbar, we normalize these physical properties.  The vertical dispersion for a given physical quantity is simply imposed jitter to clarify the location of the points.   We discuss the SHAP values in turn.

Concomitant to Figure~\ref{figure:shapley}, in
Figure~\ref{figure:2dhist}, we show 2D histograms of the physical
properties ($\rho, Z$ and $T$) of the gas in the vicinity of dust
particles, with the third dimension denoting the dust growth rate
(with an arbitrary scaling). The particles shown here are an aggregate over all snapshots from $z=14\rightarrow7$. Figure~\ref{figure:2dhist} shows the
local conditions surrounding growing dust, and aids in the interpretation of 
Figure~\ref{figure:shapley}.

Figure~\ref{figure:shapley} demonstrates the clearest physical
condition that enables the rapid growth of dust in the early Universe
is the local gas density.  Particles in high-density regions (red
points) consistently show large, positive SHAP values, signifying a
strong positive contribution to the total dust mass, while low-density
environments (blue points) strongly inhibit it.  This is made
additionally evident in the left panel of Figure~\ref{figure:2dhist}.
The positive correlation between density and dust growth is clear
(Equation~\ref{equation:growth}).  The correlation with low density
and conditions that are prohibitive for growth are also high
temperature regions: in these regions, thermal sputtering rates are high and
erode existing dust particles.   In Figure~\ref{figure:dust_dens_z}, we show a boxen plot of the distribution of gas densities in the vicinity of dust particles in our model galaxies as a function of redshift, demonstrating the rise in gas density as redshift decreases.

Second, the Shapley analysis reveals that the dust grain size
distribution has a strong correlation with dust growth, even if at the
outset challenging to interpret.  To inform this discussion, in the
left panel of Figure~\ref{figure:gsd}, we plot the redshift
evolution of the dust grain size distribution for our fiducial example galaxy.  With decreasing redshift, as the star formation rate
increases as the central galaxy in the halo grows, the dust turbulent velocity increases, and the shattering rates rise.   We demonstrate this explicitly in Figure~\ref{figure:vdisp_z}, where we show the distribution of turbulent velocities for dust particles as a function of redshift relative to our shattering thresholds. This drives a
transformation of power in the dust grain size distribution from large
grains to small grains.  This can be parameterized across all model
galaxies into a simple (but crude) statistic by collapsing the dust size spectrum  into a single value: the small-to-large ratio, defined as:
\begin{equation}
{\rm STL} \equiv \frac{M_{\rm dust} \left(\leq1.5\times10^{-3} \mu m\right)}{M_{\rm dust} \left(>1.5\times10^{-3} \mu m\right)}.
\end{equation}
We show this demarcation between
small and large in the left panel of Figure~\ref{figure:gsd},
and the redshift evolution of the small to large ratio for all
galaxies in the corresponding right panel.  While there is naturally
dispersion in the trend, there is a general sense of galaxies small to
large (STL) ratios increasing with decreasing redshift.

\begin{figure}
  \includegraphics[scale=0.35]{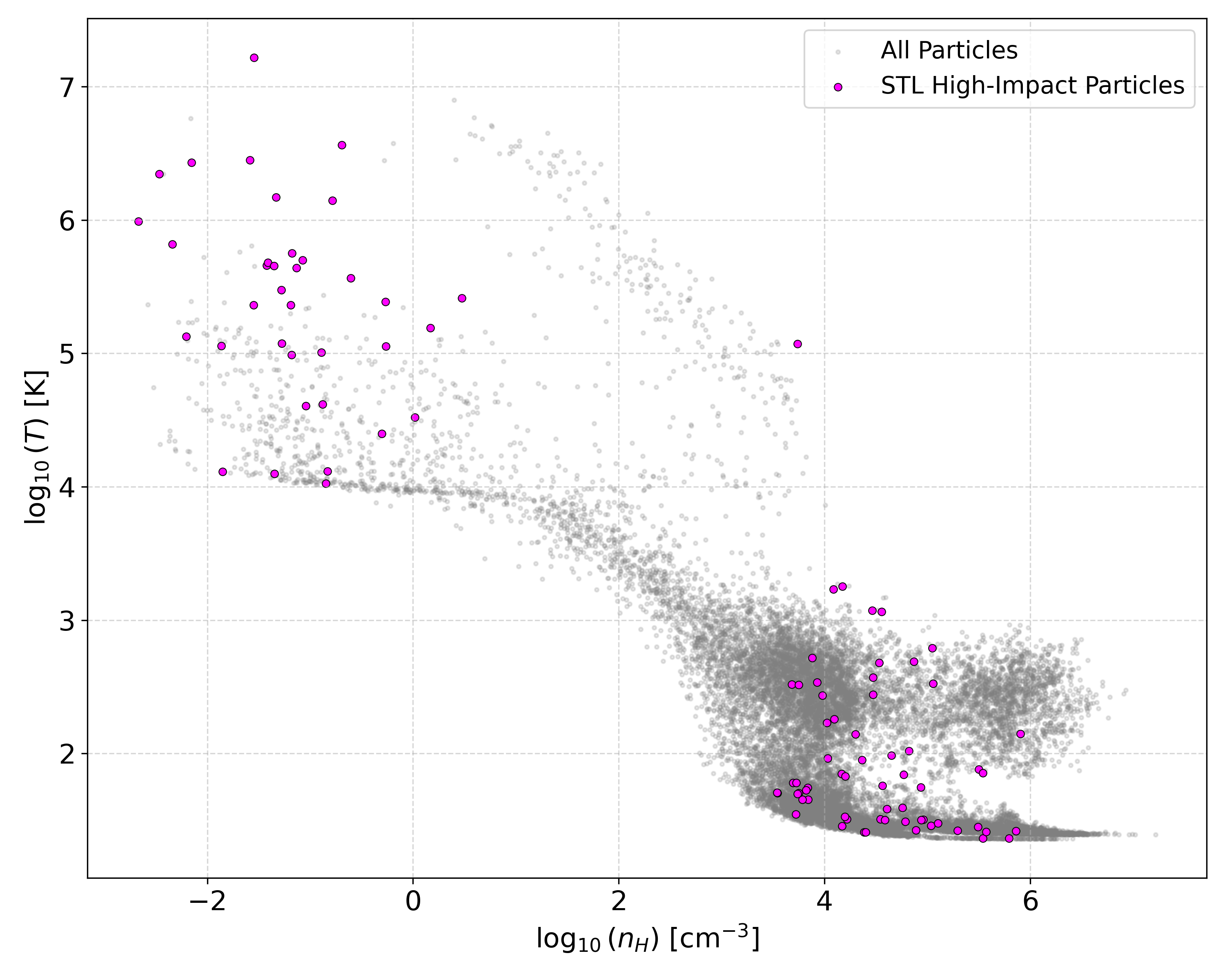}
  \caption{{\bf Dust grains whose grain size distribution has an
      outsize impact on the dust growth reside both in dense, cold
      gas, as well as diffuse, hot gas. }  This is shown in this phase
    space diagram, showing all dust particles for our fiducial example galaxy, with particles that reside in the $99^{\rm th}$
    percentile of Shapley indices (positive or negative) highlighted.
    High STL values are both conducive to dust growth in cold, dense
    gas, as well as dust sputtering in warm, diffuse gas.  
    \label{figure:rhot}}
\end{figure}

Armed with this information, we now return to the relationship between
STL ratio and dust growth in the second row of
Figure~\ref{figure:shapley}.  At face value,
Figure~\ref{figure:shapley} suggests that large STL ratios (i.e.,
grain size distributions that are heavily weighted toward small
grains) are important {\it both} for dust growth, and also correlate
with the inhibition of dust growth.  To understand this, in
Figure~\ref{figure:rhot}, we plot a $\rho-T_{\rm g}$ phase space plot
of the average gas properties in the vicinity of our model dust particles.  Recalling
Figure~\ref{figure:2dhist}, the densest gas is where the bulk of the
dust growth occurs.  We additionally highlight the model dust
particles in Figure~\ref{figure:rhot} which have an absolute value
SHAP value in the $99^{\rm th}$ percentile of the distribution of
their Shapley values: we denote these as ``STL High-Impact
Particles''.  As is immediately clear, the model dust particles with
the highest correlation with the dust growth lie both in the densest,
coldest gas, as well as the warmest and most diffuse gas.  The former
point is obvious: dust that has a size distribution
heavily skewed toward small sizes has increased surface area per unit
mass, and therefore shorter growth timescales.  To understand the
strong correlation with warm, diffuse gas, we recall that thermal
sputtering destruction timescales are shortest when dust grains are
small, and in a warm and dense environment
(Equation~\ref{equation:sputtering}).  The dust that is both smallest, and reside in the warmest gas are the most readily
thermally sputtered.  Taken together, small dust grains are both ripe
for rapid dust growth in the ISM of galaxies, as well as destruction
via thermal sputtering: the key here is the environment in which these
small dust grains reside.
When combining this with Figure~\ref{figure:gsd}, which shows
the general increase in small-to-large ratio in dust size distributions
in galaxies, the strong second-place role in the dust sizes in the
growth of dust in early Universe galaxies becomes clear.

Third, the gas temperature appears to have an inverse correlation with
the dust masses, where high temperatures are seemingly predictive of
lower dust masses, which at its face contradicts
Equation~\ref{equation:growth} (where the time scales for growth scale
inversely with $\sqrt{T_{\rm g}}$).  This said, as demonstrated in
Figure~\ref{figure:rhot}, the neighboring gas density, which is the
dominant driver in the accretion physics, is highest in cold regions.  As a result, we
conclude that the gas temperature in the vicinity of dust particles in
our simulations is subdominant in the physics of dust growth in the
early Universe.

Finally, the metallicity of the gas in the vicinity of dust has a
relatively weak correlation: indeed, generally higher metallicities
are a necessary but not solely sufficient condition for dust growth in
the ISM.  While the availability of free metals are of course
necessary for dust growth, the physical condition of the surrounding
gas (most importantly: the gas density) dominates.  High metallicity
cannot overcome the lack of growth (and, indeed, sputtering, which is
the opposite of growth) in hot, diffuse gas.  As a result, the physical correlation between gas metallicity and dust growth remain weak.

\subsection{Dust Scaling Relations in the Early Universe}

\begin{figure}
  \includegraphics[scale=0.6]{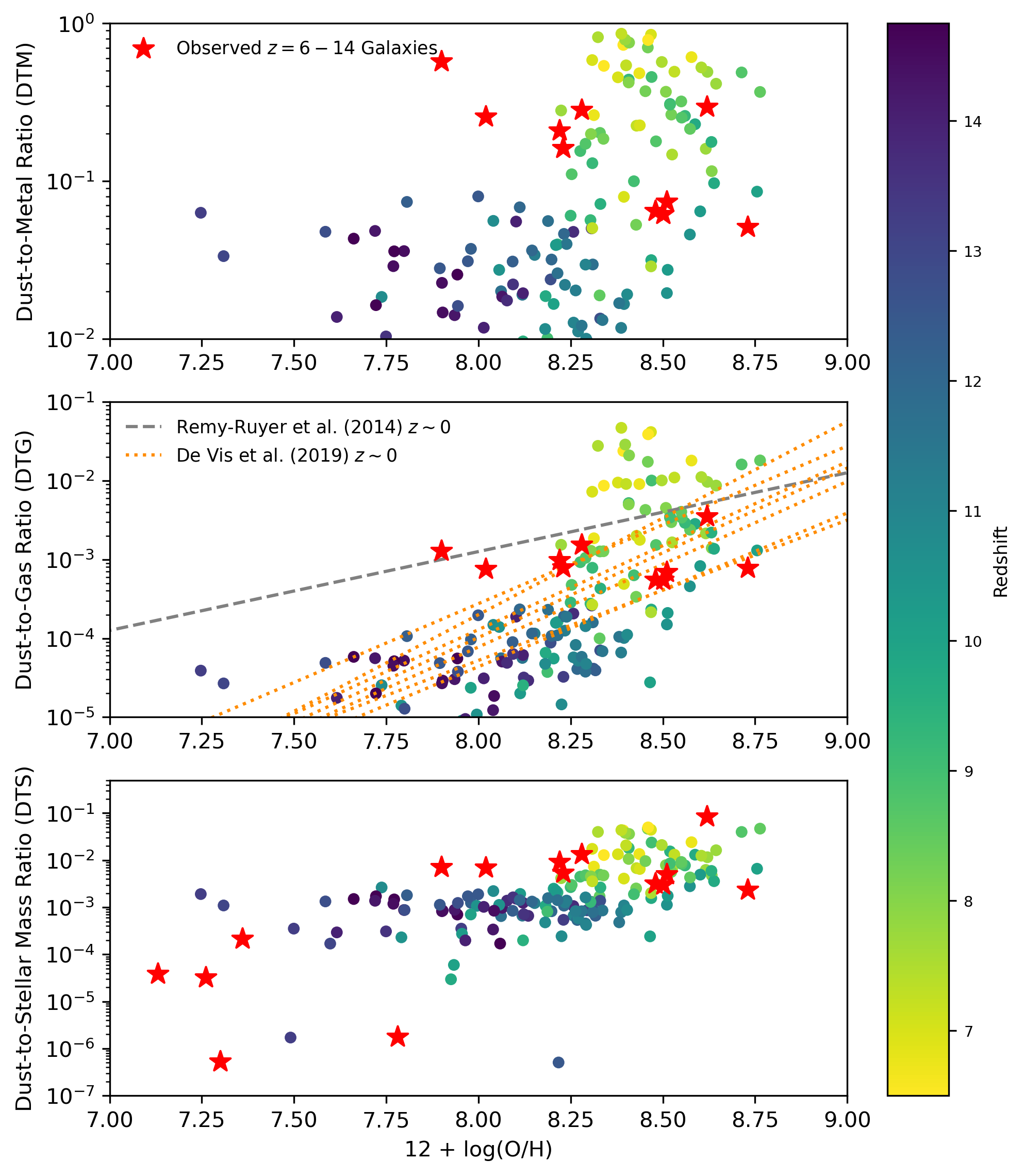}
\caption{{\bf Predicted dust scaling relations of galaxies in the
    early Universe.}.  From top to bottom, we show the dust to metal
  mass ratio (DTM), the dust to gas mass ratio (DTG), and the dust to
  stellar mass ratio (DTS), all as a function of metallicity.  In all
  plots, our model galaxies are shown by filled circles that are color
  coded by redshift, while observations are shown by red stars.  In
  the middle panel, we additionally show $z\sim0$ constraints by
  \citet{remyruyer14a} and \citet{devis19a} via dashed and dotted
  lines.  \label{figure:dust_scaling}}
\end{figure}

We now turn to the expected dust masses in the early Universe as a
function of galaxy metallicity.  In Figure~\ref{figure:dust_scaling},
we plot the dust to metal ratio (DTM), dust to gas ratio (DTG) and
dust to stellar mass ratio (DTS) all as a function of metallicity.  We
provide comparisons to both observations of $z=6-14$ galaxies \citep{ferrara22a,algera25a}
as well as fitting relations derived from low-redshift observations
\citep{remyruyer14a,devis19a}.  We discuss these relationships in turn.

The dust to metal mass and dust to gas mass ratio for our
simulated $z>7$ galaxies generally rises with metallicity, albeit with
significant scatter.  This dispersion is both intrinsic (i.e.,
snapshot to snapshot variation within the same galaxy, owing to the
regular dispersal of metals and dust in relatively shallow potential
wells in the early Universe; \citet{muratov17a}), as well as between model galaxies.  Both
the DTM and DTG ratios demonstrate a sharp uptick around $12+{\rm log}\left({\rm O/H} \right) \sim 8$,
corresponding to the approximate redshift where the model galaxies
transition from production dominated to growth dominated, and
therefore undergo a rapid increase in dust mass.

This said, while the general sense of the trend, and normalization
in the DTM and DTG ratios with metallicity are both in reasonable
agreement with lower-redshift observations, there may be some subtle tension
with inferences from observations.  \citet{algera25a} (whose data
comprises the observational data in the DTM and DTG ratio plots) find
an inverse relationship between the DTM ratio and metallicity, and a milder
increase with DTG as a function of metallicity than we predict.  However, we hesitate to
over-interpret this potential discrepancy, owing to (a) uncertainties in
single-band detections and dust mass inferences
\citep{casey12a,cochrane22a,lower24a}, as well as (b) uncertainties in the modeling procedure itself, which we discuss in \S~\ref{section:discussion}.

Finally, we turn to the modeled dust to stellar mass ratio as a
function of metallicity.  At moderate to high  metallicities, (12+log(O/H) $>8$), our simulations correspond well with the relatively few observational constraints.   It is difficult
to ascertain how well our models do in comparison to observational
inferences at low metallicities, however, owing to the relatively sparse simulation output with dust
masses at (12+log(O/H) $\la 8$).
At least some simulation points have dust to stellar ratios comparable
to those observed at 12+log(O/H)$\sim 7-8$ though some of the
observed galaxies may be in mild tension with our model predictions.


\begin{figure*}
  \centering
  \includegraphics[scale=0.45]{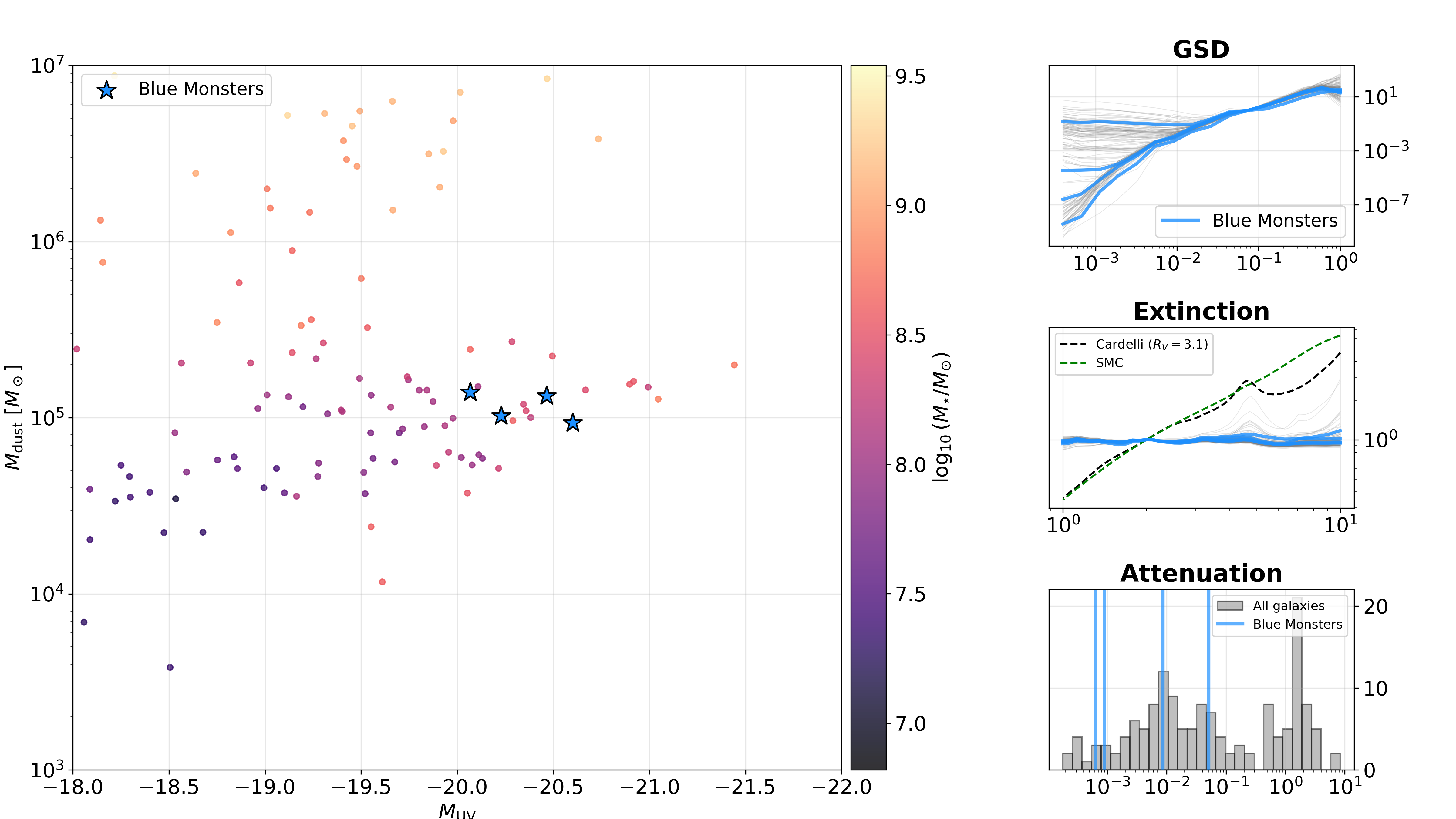}
  \caption{{\bf Blue Monsters owe their extreme UV luminosities and
      apparently lack of dust to exotic grain size distributions, and
      a resulting lack of UV opacity.}  {\it Left:} We show the
    M$_{\rm dust}$ vs the mock $M_{\rm UV}$ for all model galaxies at $z>12$, color-coded by their stellar mass.  We highlight
    Blue Monster candidates as blue stars, selected at $z>12$, with
    $M_{\rm UV} < -20$, and $M_* > 10^8$ M$_\odot$. {\it Right:} We show
    the dust grain size distributions for the same galaxies (top), the
    inferred dust extinction laws (middle),  and histogram of the $A_{\rm V}$ (bottom) from the ultimate {\it attenuation} of our blue monsters (blue), compared to the field sample (grey).  In all
    plots, we highlight the blue monster candidates in blue, to
    show its relatively top-heavy grain size distribution, its grey
    extinction curve, and its low attenuation $A_{\rm V}$.  \label{figure:blue_monsters}}
\end{figure*}

\section{The Origin of Blue Monsters}
\label{section:blue_monsters}
As demonstrated thus far, our models successfully reproduce the rapid
dust enrichment needed to explain the dusty galaxy population at $z
\sim 6-7$.  We now turn our attention to the second major topic of this paper: the origin of high-$z$ Blue Monsters.
The nature of the problem can be summarized as follows: for even the most conservative case (production-only, with no growth), we expect dust-to-stellar 
mass ratios $\sim 10^{-4}$  (Figure~\ref{figure:dust_grow}).  At the same time, JWST has
uncovered populations of galaxies at $z \gtrsim 10$ that are
exceptionally luminous in the UV ($M_{\rm UV} \la -20-21$), possess
extremely blue UV continuum slopes ($\beta \la -2.2$) and have
relatively low inferred dust to stellar mass ratios ($\leq 10^{-5}$)
\citep{ferrara23a,fiore23a,ziparo23a,ferrara24a,ferrara25a}.  We
investigate the potential origin of this population in the context of
our dust model.

Our main finding is that while none of our model galaxies
have dust to stellar mass ratios $\leq 10^{-5}$, even at $z>10$
(Figure~\ref{figure:dust_grow}), the grain size distributions at these
early times are sufficiently weighted toward large dust grains (i.e.,
top-heavy) that they are effectively transparent in the UV.  To show
this, we must convert our physical simulations of galaxy evolution
into a mock observable space in order to select potential blue monster
candidates that are sufficiently bright in the ultraviolet.  To
determine the inferred observational properties of our model, we
post-process the snapshots through {\sc powderday}
\citep{narayanan21a}, a publicly available 3D dust radiative transfer
code that uses {\sc fsps} for stellar population synthesis
\citep{conroy10a,conroy10b,conroy10c}, {\sc hyperion} for the Monte
Carlo radiative transfer \citep{robitaille11a}, and {\sc yt} for
variable and grid regularization \citep{turk11a}.  For the stellar
populations, we assume {\sc mist} stellar isochrones, and the
radiative transfer occurs on a Voronoi mesh built around the dust
particles.  The local extinction is determined by the local grain size
distribution and composition \citep{li21a}, and photons are
propagated through the dusty ISM until the equilibrium dust
temperature has converged.

In the left panel of Figure~\ref{figure:blue_monsters}, we plot the
dust masses of all model galaxies at $z>12$ vs their $M_{\rm UV}$ (measured at $\lambda = 1500 \AA$).  We
color code the model galaxies by their stellar mass, and highlight
blue monster candidates as blue stars.  We identify these candidates
as galaxies at redshift $z>12$, $M_{\rm UV} < -20$, and $M_* > 10^8
M_\odot$.  Right away, it is clear that massive, early, and extremely
UV luminous systems exist in our simulations.

The blue monster candidates in our model galaxies have high UV
luminosities owing (on average) to their relatively top-heavy weighted
grain size distributions.  Recalling \S~\ref{section:methods}, when
dust is ejected from SNe, it has a lognormal size distribution
centered at $a_0 = 0.1 \mu$m owing to the destruction of smaller
grains in the reverse shock.  As the galaxy assembles and the velocity
dispersion of our model dust particles increases, the shattering rates
increase, and the size distribution shifts toward smaller sizes
(c.f. Figure~\ref{figure:gsd}).  However, at $z>10$, while our model
massive galaxies indeed have substantial dust to stellar mass ratios,
the grain size distributions have not yet had time to shift their
power toward small sizes via grain-grain shattering, and therefore are
relatively optically thin to UV radiation.  This is seen explicitly in
the top right of Figure~\ref{figure:blue_monsters}, where we show the
grain size distribution of all of our model galaxies presented in the
$M_*-M_{\rm UV}$ plane, and in the bottom panel where we show the
corresponding extinction law.  We highlight in blue the blue monster
candidates.  The top heavy grain size distribution of our model blue
monsters results in relatively grey extinction laws, rendering the
model blue monsters optically thin in the UV.  Ultimately, there
  is one additional step in the UV observability of these galaxies,
  which is the radiative transfer through complex star-dust
  geometries.  In the bottom-panel of
  Figure~\ref{figure:blue_monsters}, we confirm that these galaxies
  have sufficiently grey attenuation laws that the ultimate $A_{\rm
    V}$ is relatively low for our blue monsters (blue), compared to
  the parent population of galaxies (grey).\footnote{The star-dust geometry/attenuation effects result in even some galaxies with larger abundances of small dust grains to appear as blue monsters.  One such example exists in Figure~\ref{figure:blue_monsters}.}

In summary, Figures~\ref{figure:gsd}, ~\ref{figure:dust_grow} and
~\ref{figure:blue_monsters} paint a picture in which massive galaxies
at $z>10$ have dust to stellar mass ratios as expected from
production+growth scenarios ($\sim 10^{-4}-10^{-3}$), their grain size
distributions have not yet shifted sufficient power to small grains.
As a result, we conclude that blue monsters are likely simply galaxies
which are optically thin owing to their grain size distributions.  We
discuss the consequences of this further in
\S~\ref{section:discussion_blue_monsters}.

\section{Discussion}
\label{section:discussion}
\subsection{The Growth of the First Dust: Theoretical Perspective}

We begin our discussion by comparing our results to existing results
in the literature that have attempted to model the formation of the
first dust.  Indeed, quite a number of studies in the literature
already have investigated the growth of dust in the early Universe
\citep[e.g.][]{vijayan19a,graziani20a,esmerian22a,vijayan22a,esmerian23a,dicesare23a,lewis23a,lower23a,choban24a,jones24a,lower24a,choban25a,trayford25a}.
What distinguishes the model presented in this work is the combination both of an explicitly
resolved ISM model, and --critically-- a multi-size live dust
population.  We find that the dominant contributors to the dust growth
in the early Universe are the buildup of dense gas near dust
particles, as well as a shifting in the power of the grain size
distribution toward smaller grains.  This leads to a natural question:
the aforementioned simulations by and large claim some level of
success in matching dust mass constraints from reionization-epoch
galaxies, albeit with fairly different dust implementations and even
more broadly, underlying galaxy evolution physics.  At the same time,
some other studies \citep[e.g.][]{choban25a} have found challenges in
matching the dust content of some of the earliest galaxies -- a result
that is not entirely clear how to mesh with our own, given the
relative similarities between the {\sc fire} model that underpins the
\citet{choban25a} simulations, and the {\sc smuggle} physics in our
own.  How can these seemingly disparate results be reconciled?

\citet{esmerian22a,esmerian23a}  may offer a clue.  In
particular, \citet{esmerian22a} performed a parameter exploration of
the {\sc croc} reionization-epoch simulations, ultimately finding that
the total dust mass is sensitive to the underlying choice of dust
physical parameters.  These include, for example, evolved star dust
yields, characteristic growth time scales, parameters associated with
destruction efficiency, and subresolution clumping factors.  The
variations investigated in \citet{esmerian22a} explore ranges in
parameter space that are reasonable given the general uncertainty in
our understanding of fundamental dust physics in the ISM of galaxies.

At the same time, beyond the inherent uncertainty in subresolution
dust parameters\footnote{As a quantitative example, we assume
accretion time scales $\sim 0.2$ Gyr, while \citet{esmerian23a} assume
a value $\sim 50\%$ larger; both are reasonable choices within
traditionally accepted ranges \citep{li18a}, and yet will result in
different growth histories.}, there is a complex interplay between the
implemented dust physics, and the underlying galaxy formation physics.
For example, the dust content and growth history will impact the
available free metals, which impacts cooling, which impacts star
formation, which impacts future dust production, and so on.  Moreover,
feedback physics will impact the gas density distribution function,
gas temperature, and star formation history (which of course impacts
the amount of free metals).  The interdependencies are complex and
often nonlinear.  As a result, the exact details of the physical
properties of the gas in the vicinity of growing dust will vary
dramatically from simulation to simulation, even for simulations that
include reasonably similar underlying physics \citep[e.g. our model
  and that of ][]{choban25a}.  For example, in our work, we
  employ a modest star formation threshold of $n_{\rm thresh} = 150$
  cm$^{-3}$, while \citet{choban25a} employ a threshold density
  $n_{\rm thresh} = 1000$ cm$^{-3}$.  This choice may result in
  smoother star formation histories in our model than the {\sc fire-2}
  model, as well as more steady moderate density gas as compared to
  larger swings in density (owing to the collapse-feedback
  ``breathing'' mode of star formation).  This may result in more
  steady accretion onto dust grains in our model compared to the
  \citet{choban25a} model.


The upshot is that there are  degenerate solutions in the
pantheon of possible parameter choices for any given model that will
simultaneously reproduce both basic galaxy scaling relations that
cosmological simulations are often tuned to \citep[e.g. the mass-metallicity
relation, stellar mass function, stellar mass-halo mass relation,
etc.;][]{ni23a,garcia24a,garcia25a}, and also standard dust scaling relations.  Absent a side-by-side code comparison project starting from the same sets of initial conditions, these degeneracies in modeling make it challenging to disentangle exactly why some simulations are able to match the inferred dust masses in the Epoch of Reionization, while others are not. While our model is
the first to include the physics of a grain size distribution on the
fly -- and find that the growth history depends on this grain size
distribution -- it is easy to envision cases which adopt other
reasonable parameter choices that obviate the role of the grain size
distribution in the dust growth history of a galaxy.   Fundamentally, galaxy evolution
simulations should be considered as numerical experiments, and the
results presented from those, including our own, are simply the
consequence of an adopted physical implementation within that
experiment.

\subsection{Blue Monsters and the Attenuation Free Model}
\label{section:discussion_blue_monsters}
Our model suggests that the origin of blue monsters, the population of
$z \ga 10$ extremely UV bright and potentially massive galaxies that
show little evidence for dust reddening actually has an expected
dust-to-stellar mass ratio, but fairly non-standard dust grain size
distributions.  To put it more simply: the dust is there, but we just cannot
see it.

The model that we present is in subtle contrast to the leading theory
employed to understand the origin of these galaxies: the attenuation
free model
\citep[AFM;][]{ferrara23a,ziparo23a,ferrara24a,ferrara25b,ferrara25a}.
In the AFM model, dust is produced normally at even higher
redshifts (say, $z>14$), but then launched by stellar winds into an
extended geometry, therefore reducing the effective optical depth.
The AFM class of models, significantly better studied in the
literature than our own, has a demonstrated success of matching the
relatively slow evolution of the high-$z$ UV luminosity function
\citep{ferrara23a}, the high-$z$ star formation rate density
evolution, the observed Ly$\alpha$ properties of some $z>10$ galaxies
\citep{ferrara24c}, and the inferred star formation histories of some
individual galaxies \citep{ferrara24b}.  In their semi-analytic
models, \citet{zhao24a} similarly inferred a need for extended dust
distributions in $z>5$ galaxies in order to simultaneously match the
observed UV and infrared properties of galaxies in this epoch.

While at their core, the AFM model and our model are fairly different,
they share a similar fundamental ethos: both models agree that the
dust has always been there, as expected, and that the lack of
reddening in $z>10$ UV luminous galaxies is just due to a lack of
attenuation.  In that sense, our model too could be considered as a
member of the class of AFM models, with the key difference being {\it
  how} the attenuation is reduced.  In our picture, the lack of
attenuation in the rest frame UV arises from grain size distribution
itself.  The two models may possibly be distinguished via searches for
infrared detection in galaxies at $z\ga z_{\rm outflow}$, where
$z_{\rm outflow}$ is the typical redshift of dusty outflows in the
traditional \citet{ferrara23a} AFM outflow model.  At redshifts above
$z_{\rm outflow}$, the traditional AFM model may predict dust obscured
star formation (and hence, infrared emission), whereas our model which
is large-grain dominated at these redshifts would imply almost no
reprocessing of UV photons into the infrared at these redshifts.  A
more thorough exploration of model-testing predictions will be
presented in a future paper.

Our model, which is predicated on the small to large fraction in dust
sizes increasing with decreasing redshift, results in 3 key corollary
consequences, all of which are deferred to planned ongoing studies.
First, the extinction laws will go from grey to steep as the fraction
of small grains rises with cosmic time.  Absent complications with the
star-dust geometry, this will likely leave an imprint on the
galaxy-wide attenuation law as well \citep{salim20a}.  Indeed,
\citet{markov24a} have detected this potential signature in the
inferred evolution of dust attenuation laws from $z \sim 2\rightarrow
12$, and this may be expected from theoretical arguments \citep{mckinney25a}.  Second, an increase in grain-grain shattering rates, and
ultrasmall grain fractions, may result in the emergence of the $2175
\AA$ bump in extinction/attenuation.  Some tentative
observational evidence for this feature has appeared in JWST data
\citep{witstok23a,markov24a,ormerod25a}.  Already we can start to see
evidence for the UV bump in our model extinction laws in Figure~\ref{figure:blue_monsters}.  Both
the evolution of the attenuation law at high-$z$ and emergence of the
$2175\AA$ bump will be explored in forthcoming work (D. Zimmerman et
al., in prep, and E. Savitch et al. in prep.).

Finally, one of the strongest predictions made by the original,
outflows-based AFM model, is that it naturally results in a relatively
slow evolution of the $z>8$ UV luminosity function, a trend that has
been well documented in numerous observational studies
\citep[e.g.][and references
  therein]{arrabalharo23a,curtislake23a,finkelstein23a,harikane23a,robertson23a,donnan23a,donnan23b}.
At face value, it is likely that our model will at least help boost
the UV luminosity per galaxy stellar mass at $z>8$, owing to reduced
obscuration, and therefore offsetting (at least somewhat) the drop in abundance of massive galaxies with increasing redshift.  A forthcoming paper (J. Kelley-Derzon et al. in prep.) will explore the consequences of our dust model on the evolution of the high-$z$ UVLF.

\section{Summary}
\label{section:summary}

In this paper, we have combined cosmological zoom-in simulations of
massive early Universe galaxies with a comprehensive dust evolution
model to investigate the rapid buildup of dust in the first billion
years and the origin of blue monster galaxies at $z>10$. We have designed
our study as a series of numerical experiments to isolate the key
physical drivers of early dust growth.  Our main results follow.

\begin{enumerate}
\item {\bf Dust production dominates the dust content of galaxies through
  $z\approx 10-11$ for galaxies in our modeled mass range ($M_*
  \approx 10^9-10^{10} M_\odot$ at $z=7$), at which point growth from
  metal accretion in the ISM takes over.}  Production only models reach
  $M_{\rm dust}/M_* \approx 10^{-4}$, though a growth dominated regime
  is necessary to reach the observed ratios of $M_{\rm dust}/M_* \approx 10^{-3}-10^{-2}$ by $z \approx 6$, consistent with ALMA detections (Figure~\ref{figure:dust_grow}).
  
\item {\bf The local gas density is the primary driver of dust growth
  in early Universe galaxies.}  We have demonstrated this via Shapley
  feature importance analysis.  The
  hierarchical assembly of galaxies between $z=14\rightarrow7$ leads
  to dramatic increase in gas density near dust particles, enabling
  rapid metal accretion (Figure~\ref{figure:shapley}).    

\item {\bf The grain size distribution is the second most important factor driving dust growth in early Universe galaxies.}  Smaller grains grow faster owing to increased surface area per unit mass.   Grain-grain shattering drives a rapid shift in the grain size distribution toward smaller dust grains with decreasing redshift.  The small-to-large grain ratio increases by factors of $\sim 10-100$ between $z=12\rightarrow7$ (Figure~\ref{figure:gsd}).

       \item {\bf The modeled dust scaling relations broadly match
        observations at $z \sim 6-7$} with dust-to-metal and
        dust-to-gas ratios rising with metallicity, though with
        significant intrinsic scatter owing to the bursty nature of
        early star formation. The transition from production-dominated
        to growth-dominated regimes occurs around 12+log(O/H) $\sim 8$ (Figure~\ref{figure:dust_scaling}). 

    \item {\bf Blue Monsters arise naturally from top-heavy grain size
      distributions, rather than exotic dust geometries.}  In our
      simulations, the population of massive, UV-bright galaxies
      discovered at $z>10$ have dust-to-stellar mass ratios consistent with production-dominated scenarios (i.e., $\sim
      10^{-4}$).  This said, they are detected before significant
      grain-grain shattering has had time to occur, and their grain
      size distributions are weighted toward large grains, rendering
      these galaxies optically thin in the UV despite harboring
      substantial dust reservoirs (Figure~\ref{figure:blue_monsters}).

\end{enumerate}

\section{Acknowledgements}
The authors thank the anonymous referee for their constructive
reports, which we feel improved the paper. D.N. is grateful to Caleb
Choban, Andrea Ferrara, Jed McKinney, Massimiliano Parente, Gergo
Popping, Rafaella Schneider \& Rachel Somerville for helpful
conversations.  D.N. would also like to express thanks to Adriano
Fontana, Paola Santini, Jim Dunlop, Davide Elbaz, Alice Shapley and
Rachel Somerville for their work in organizing the conference ``The
Growth of Galaxies in the Early Universe - X'' at the Sexten Center
for Astrophysics, January 2025, during which the main ideas for this
paper were generated.  D.N. is grateful to Shazrene Mohamed and the
University of Virginia Department of Astronomy for hosting him during
the bulk of paper writing.  D.N. and P.T. were funded by NASA
ATP21-0013, ``Direct Modeling of Interstellar Dust in a Cosmological
Framework.''



\begin{thebibliography}{}
\makeatletter
\relax
\def\mn@urlcharsother{\let\do\@makeother \do\$\do\&\do\#\do\^\do\_\do\%\do\~}
\def\mn@doi{\begingroup\mn@urlcharsother \@ifnextchar [ {\mn@doi@}
  {\mn@doi@[]}}
\def\mn@doi@[#1]#2{\def\@tempa{#1}\ifx\@tempa\@empty \href
  {http://dx.doi.org/#2} {doi:#2}\else \href {http://dx.doi.org/#2} {#1}\fi
  \endgroup}
\def\mn@eprint#1#2{\mn@eprint@#1:#2::\@nil}
\def\mn@eprint@arXiv#1{\href {http://arxiv.org/abs/#1} {{\tt arXiv:#1}}}
\def\mn@eprint@dblp#1{\href {http://dblp.uni-trier.de/rec/bibtex/#1.xml}
  {dblp:#1}}
\def\mn@eprint@#1:#2:#3:#4\@nil{\def\@tempa {#1}\def\@tempb {#2}\def\@tempc
  {#3}\ifx \@tempc \@empty \let \@tempc \@tempb \let \@tempb \@tempa \fi \ifx
  \@tempb \@empty \def\@tempb {arXiv}\fi \@ifundefined
  {mn@eprint@\@tempb}{\@tempb:\@tempc}{\expandafter \expandafter \csname
  mn@eprint@\@tempb\endcsname \expandafter{\@tempc}}}

\bibitem[\protect\citeauthoryear{{Algera} et~al.,}{{Algera}
  et~al.}{2024}]{algera24a}
{Algera} H. S.~B.,  et~al., 2024, \mn@doi [\mnras] {10.1093/mnras/stae1994},
  \href {https://ui.adsabs.harvard.edu/abs/2024MNRAS.533.3098A} {533, 3098}

\bibitem[\protect\citeauthoryear{{Algera} et~al.,}{{Algera}
  et~al.}{2025}]{algera25a}
{Algera} H.,  et~al., 2025, \mn@doi [arXiv/2501.10508]
  {10.48550/arXiv.2501.10508}, \href
  {https://ui.adsabs.harvard.edu/abs/2025arXiv250110508A} {p. arXiv:2501.10508}

\bibitem[\protect\citeauthoryear{{Arrabal Haro} et~al.,}{{Arrabal Haro}
  et~al.}{2023}]{arrabalharo23a}
{Arrabal Haro} P.,  et~al., 2023, \mn@doi [\apjl] {10.3847/2041-8213/acdd54},
  \href {https://ui.adsabs.harvard.edu/abs/2023ApJ...951L..22A} {951, L22}

\bibitem[\protect\citeauthoryear{{Asada} et~al.,}{{Asada}
  et~al.}{2024}]{asada24a}
{Asada} Y.,  et~al., 2024, \mn@doi [\mnras] {10.1093/mnras/stad3902}, \href
  {https://ui.adsabs.harvard.edu/abs/2024MNRAS.52711372A} {527, 11372}

\bibitem[\protect\citeauthoryear{{Asano}, {Takeuchi}, {Hirashita}  \&
  {Inoue}}{{Asano} et~al.}{2013}]{asano13a}
{Asano} R.~S.,  {Takeuchi} T.~T.,  {Hirashita} H.,   {Inoue} A.~K.,  2013,
  \mn@doi [Earth, Planets, and Space] {10.5047/eps.2012.04.014}, \href
  {https://ui.adsabs.harvard.edu/abs/2013EP&S...65..213A} {65, 213}

\bibitem[\protect\citeauthoryear{{Austin} et~al.,}{{Austin}
  et~al.}{2023}]{austin23a}
{Austin} D.,  et~al., 2023, \mn@doi [\apjl] {10.3847/2041-8213/ace18d}, \href
  {https://ui.adsabs.harvard.edu/abs/2023ApJ...952L...7A} {952, L7}

\bibitem[\protect\citeauthoryear{{Austin} et~al.,}{{Austin}
  et~al.}{2024}]{austin24a}
{Austin} D.,  et~al., 2024, \mn@doi [arXiv/2404.10751]
  {10.48550/arXiv.2404.10751}, \href
  {https://ui.adsabs.harvard.edu/abs/2024arXiv240410751A} {p. arXiv:2404.10751}

\bibitem[\protect\citeauthoryear{{Bakx} et~al.,}{{Bakx} et~al.}{2021}]{bakx21a}
{Bakx} T. J.~L.~C.,  et~al., 2021, \mn@doi [\mnras] {10.1093/mnrasl/slab104},
  \href {https://ui.adsabs.harvard.edu/abs/2021MNRAS.508L..58B} {508, L58}

\bibitem[\protect\citeauthoryear{{Barlow} et~al.,}{{Barlow}
  et~al.}{2010}]{barlow10a}
{Barlow} M.~J.,  et~al., 2010, \mn@doi [\aap] {10.1051/0004-6361/201014585},
  \href {https://ui.adsabs.harvard.edu/abs/2010A&A...518L.138B} {518, L138}

\bibitem[\protect\citeauthoryear{{Bassini}, {Feldmann}, {Gensior},
  {Faucher-Gigu{\`e}re}, {Cenci}, {Moreno}, {Bernardini}  \& {Liang}}{{Bassini}
  et~al.}{2024}]{bassini24a}
{Bassini} L.,  {Feldmann} R.,  {Gensior} J.,  {Faucher-Gigu{\`e}re} C.-A.,
  {Cenci} E.,  {Moreno} J.,  {Bernardini} M.,   {Liang} L.,  2024, \mn@doi
  [\mnras] {10.1093/mnrasl/slae036}, \href
  {https://ui.adsabs.harvard.edu/abs/2024MNRAS.532L..14B} {532, L14}

\bibitem[\protect\citeauthoryear{{Behroozi}, {Wechsler}  \&
  {Conroy}}{{Behroozi} et~al.}{2013}]{behroozi13a}
{Behroozi} P.~S.,  {Wechsler} R.~H.,   {Conroy} C.,  2013, \mn@doi [\apj]
  {10.1088/0004-637X/770/1/57}, \href
  {https://ui.adsabs.harvard.edu/abs/2013ApJ...770...57B} {770, 57}

\bibitem[\protect\citeauthoryear{{Bianchi} \& {Schneider}}{{Bianchi} \&
  {Schneider}}{2007}]{bianchi07a}
{Bianchi} S.,  {Schneider} R.,  2007, \mn@doi [\mnras]
  {10.1111/j.1365-2966.2007.11829.x}, \href
  {https://ui.adsabs.harvard.edu/abs/2007MNRAS.378..973B} {378, 973}

\bibitem[\protect\citeauthoryear{{Bouwens} et~al.,}{{Bouwens}
  et~al.}{2010}]{bouwens10a}
{Bouwens} R.~J.,  et~al., 2010, \mn@doi [\apjl] {10.1088/2041-8205/708/2/L69},
  \href {https://ui.adsabs.harvard.edu/abs/2010ApJ...708L..69B} {708, L69}

\bibitem[\protect\citeauthoryear{{Bouwens} et~al.,}{{Bouwens}
  et~al.}{2012}]{bouwens12a}
{Bouwens} R.~J.,  et~al., 2012, \mn@doi [\apj] {10.1088/0004-637X/754/2/83},
  \href {https://ui.adsabs.harvard.edu/abs/2012ApJ...754...83B} {754, 83}

\bibitem[\protect\citeauthoryear{{Bouwens} et~al.,}{{Bouwens}
  et~al.}{2016}]{bouwens16a}
{Bouwens} R.,  et~al., 2016, arXiv/1606.05280, \href
  {http://adsabs.harvard.edu/abs/2016arXiv160605280B} {}

\bibitem[\protect\citeauthoryear{{Bunker} et~al.,}{{Bunker}
  et~al.}{2024}]{bunker24a}
{Bunker} A.~J.,  et~al., 2024, \mn@doi [\aap] {10.1051/0004-6361/202347094},
  \href {https://ui.adsabs.harvard.edu/abs/2024A&A...690A.288B} {690, A288}

\bibitem[\protect\citeauthoryear{{Burgarella} et~al.,}{{Burgarella}
  et~al.}{2025}]{burgarella25a}
{Burgarella} D.,  et~al., 2025, \mn@doi [\aap] {10.1051/0004-6361/202554231},
  \href {https://ui.adsabs.harvard.edu/abs/2025A&A...699A.336B} {699, A336}

\bibitem[\protect\citeauthoryear{{Byler}, {Dalcanton}, {Conroy}  \&
  {Johnson}}{{Byler} et~al.}{2017}]{byler17a}
{Byler} N.,  {Dalcanton} J.~J.,  {Conroy} C.,   {Johnson} B.~D.,  2017, \mn@doi
  [\apj] {10.3847/1538-4357/aa6c66}, \href
  {https://ui.adsabs.harvard.edu/abs/2017ApJ...840...44B} {840, 44}

\bibitem[\protect\citeauthoryear{{Calabr{\`o}} et~al.,}{{Calabr{\`o}}
  et~al.}{2021}]{calabro21a}
{Calabr{\`o}} A.,  et~al., 2021, \mn@doi [\aap] {10.1051/0004-6361/202039244},
  \href {https://ui.adsabs.harvard.edu/abs/2021A&A...646A..39C} {646, A39}

\bibitem[\protect\citeauthoryear{{Calzetti}}{{Calzetti}}{1997}]{calzetti97a}
{Calzetti} D.,  1997, \mn@doi [\aj] {10.1086/118242}, \href
  {http://adsabs.harvard.edu/abs/1997AJ....113..162C} {113, 162}

\bibitem[\protect\citeauthoryear{{Calzetti}, {Kinney}  \&
  {Storchi-Bergmann}}{{Calzetti} et~al.}{1994}]{calzetti94a}
{Calzetti} D.,  {Kinney} A.~L.,   {Storchi-Bergmann} T.,  1994, \mn@doi [\apj]
  {10.1086/174346}, \href {http://adsabs.harvard.edu/abs/1994ApJ...429..582C}
  {429, 582}

\bibitem[\protect\citeauthoryear{{Casey}}{{Casey}}{2012}]{casey12a}
{Casey} C.~M.,  2012, \mn@doi [\mnras] {10.1111/j.1365-2966.2012.21455.x},
  \href {http://adsabs.harvard.edu/abs/2012MNRAS.425.3094C} {425, 3094}

\bibitem[\protect\citeauthoryear{{Chakraborty} et~al.,}{{Chakraborty}
  et~al.}{2025}]{chakraborty25a}
{Chakraborty} P.,  et~al., 2025, \mn@doi [\apj] {10.3847/1538-4357/adc7b5},
  \href {https://ui.adsabs.harvard.edu/abs/2025ApJ...985...24C} {985, 24}

\bibitem[\protect\citeauthoryear{{Chemerynska} et~al.,}{{Chemerynska}
  et~al.}{2024}]{chemerynska24a}
{Chemerynska} I.,  et~al., 2024, \mn@doi [\apjl] {10.3847/2041-8213/ad8dc9},
  \href {https://ui.adsabs.harvard.edu/abs/2024ApJ...976L..15C} {976, L15}

\bibitem[\protect\citeauthoryear{{Cherchneff} \& {Dwek}}{{Cherchneff} \&
  {Dwek}}{2010}]{cherchneff10a}
{Cherchneff} I.,  {Dwek} E.,  2010, \mn@doi [\apj] {10.1088/0004-637X/713/1/1},
  \href {https://ui.adsabs.harvard.edu/abs/2010ApJ...713....1C} {713, 1}

\bibitem[\protect\citeauthoryear{{Choban}, {Kere{\v{s}}}, {Hopkins},
  {Sandstrom}, {Hayward}  \& {Faucher-Gigu{\`e}re}}{{Choban}
  et~al.}{2022}]{choban22a}
{Choban} C.~R.,  {Kere{\v{s}}} D.,  {Hopkins} P.~F.,  {Sandstrom} K.~M.,
  {Hayward} C.~C.,   {Faucher-Gigu{\`e}re} C.-A.,  2022, \mn@doi [\mnras]
  {10.1093/mnras/stac1542}, \href
  {https://ui.adsabs.harvard.edu/abs/2022MNRAS.514.4506C} {514, 4506}

\bibitem[\protect\citeauthoryear{{Choban}, {Kere{\v{s}}}, {Sandstrom},
  {Hopkins}, {Hayward}  \& {Faucher-Gigu{\`e}re}}{{Choban}
  et~al.}{2024}]{choban24a}
{Choban} C.~R.,  {Kere{\v{s}}} D.,  {Sandstrom} K.~M.,  {Hopkins} P.~F.,
  {Hayward} C.~C.,   {Faucher-Gigu{\`e}re} C.-A.,  2024, \mn@doi [\mnras]
  {10.1093/mnras/stae716}, \href
  {https://ui.adsabs.harvard.edu/abs/2024MNRAS.529.2356C} {529, 2356}

\bibitem[\protect\citeauthoryear{{Choban}, {Salim}, {Kere{\v{s}}}, {Hayward}
  \& {Sandstrom}}{{Choban} et~al.}{2025}]{choban25a}
{Choban} C.~R.,  {Salim} S.,  {Kere{\v{s}}} D.,  {Hayward} C.~C.,   {Sandstrom}
  K.~M.,  2025, \mn@doi [\mnras] {10.1093/mnras/staf118}, \href
  {https://ui.adsabs.harvard.edu/abs/2025MNRAS.537.1518C} {537, 1518}

\bibitem[\protect\citeauthoryear{{Ciesla} et~al.,}{{Ciesla}
  et~al.}{2024}]{ciesla24a}
{Ciesla} L.,  et~al., 2024, \mn@doi [\aap] {10.1051/0004-6361/202348091}, \href
  {https://ui.adsabs.harvard.edu/abs/2024A&A...686A.128C} {686, A128}

\bibitem[\protect\citeauthoryear{{Clarke}, {Shapley}, {Sanders}, {Topping},
  {Brammer}, {Bento}, {Reddy}  \& {Kehoe}}{{Clarke} et~al.}{2024}]{clarke24a}
{Clarke} L.,  {Shapley} A.~E.,  {Sanders} R.~L.,  {Topping} M.~W.,  {Brammer}
  G.~B.,  {Bento} T.,  {Reddy} N.~A.,   {Kehoe} E.,  2024, \mn@doi [\apj]
  {10.3847/1538-4357/ad8ba4}, \href
  {https://ui.adsabs.harvard.edu/abs/2024ApJ...977..133C} {977, 133}

\bibitem[\protect\citeauthoryear{{Cochrane}, {Hayward}  \&
  {Angl{\'e}s-Alc{\'a}zar}}{{Cochrane} et~al.}{2022}]{cochrane22a}
{Cochrane} R.~K.,  {Hayward} C.~C.,   {Angl{\'e}s-Alc{\'a}zar} D.,  2022,
  \mn@doi [\apjl] {10.3847/2041-8213/ac951d}, \href
  {https://ui.adsabs.harvard.edu/abs/2022ApJ...939L..27C} {939, L27}

\bibitem[\protect\citeauthoryear{{Cochrane}, {Katz}, {Begley}, {Hayward}  \&
  {Best}}{{Cochrane} et~al.}{2025}]{cochrane25a}
{Cochrane} R.~K.,  {Katz} H.,  {Begley} R.,  {Hayward} C.~C.,   {Best} P.~N.,
  2025, \mn@doi [\apjl] {10.3847/2041-8213/ad9a4d}, \href
  {https://ui.adsabs.harvard.edu/abs/2025ApJ...978L..42C} {978, L42}

\bibitem[\protect\citeauthoryear{{Conroy} \& {Gunn}}{{Conroy} \&
  {Gunn}}{2010}]{conroy10b}
{Conroy} C.,  {Gunn} J.~E.,  2010, \mn@doi [\apj]
  {10.1088/0004-637X/712/2/833}, \href
  {http://adsabs.harvard.edu/abs/2010ApJ...712..833C} {712, 833}

\bibitem[\protect\citeauthoryear{{Conroy}, {White}  \& {Gunn}}{{Conroy}
  et~al.}{2010a}]{conroy10a}
{Conroy} C.,  {White} M.,   {Gunn} J.~E.,  2010a, \mn@doi [\apj]
  {10.1088/0004-637X/708/1/58}, \href
  {http://adsabs.harvard.edu/abs/2010ApJ...708...58C} {708, 58}

\bibitem[\protect\citeauthoryear{{Conroy}, {Schiminovich}  \&
  {Blanton}}{{Conroy} et~al.}{2010b}]{conroy10c}
{Conroy} C.,  {Schiminovich} D.,   {Blanton} M.~R.,  2010b, \mn@doi [\apj]
  {10.1088/0004-637X/718/1/184}, \href
  {http://adsabs.harvard.edu/abs/2010ApJ...718..184C} {718, 184}

\bibitem[\protect\citeauthoryear{{Cooray} et~al.,}{{Cooray}
  et~al.}{2014}]{cooray14a}
{Cooray} A.,  et~al., 2014, \mn@doi [\apj] {10.1088/0004-637X/790/1/40}, \href
  {https://ui.adsabs.harvard.edu/abs/2014ApJ...790...40C} {790, 40}

\bibitem[\protect\citeauthoryear{{Cullen} et~al.,}{{Cullen}
  et~al.}{2024}]{cullen24a}
{Cullen} F.,  et~al., 2024, \mn@doi [\mnras] {10.1093/mnras/stae1211}, \href
  {https://ui.adsabs.harvard.edu/abs/2024MNRAS.531..997C} {531, 997}

\bibitem[\protect\citeauthoryear{{Curtis-Lake} et~al.,}{{Curtis-Lake}
  et~al.}{2023}]{curtislake23a}
{Curtis-Lake} E.,  et~al., 2023, \mn@doi [Nature Astronomy]
  {10.1038/s41550-023-01918-w}, \href
  {https://ui.adsabs.harvard.edu/abs/2023NatAs...7..622C} {7, 622}

\bibitem[\protect\citeauthoryear{{Dayal} et~al.,}{{Dayal}
  et~al.}{2022}]{dayal22a}
{Dayal} P.,  et~al., 2022, \mn@doi [\mnras] {10.1093/mnras/stac537}, \href
  {https://ui.adsabs.harvard.edu/abs/2022MNRAS.512..989D} {512, 989}

\bibitem[\protect\citeauthoryear{{De Looze}, {Barlow}, {Swinyard}, {Rho},
  {Gomez}, {Matsuura}  \& {Wesson}}{{De Looze} et~al.}{2017}]{delooze17a}
{De Looze} I.,  {Barlow} M.~J.,  {Swinyard} B.~M.,  {Rho} J.,  {Gomez} H.~L.,
  {Matsuura} M.,   {Wesson} R.,  2017, \mn@doi [\mnras]
  {10.1093/mnras/stw2837}, \href
  {https://ui.adsabs.harvard.edu/abs/2017MNRAS.465.3309D} {465, 3309}

\bibitem[\protect\citeauthoryear{{De Vis} et~al.,}{{De Vis}
  et~al.}{2019}]{devis19a}
{De Vis} P.,  et~al., 2019, \mn@doi [\aap] {10.1051/0004-6361/201834444}, \href
  {https://ui.adsabs.harvard.edu/abs/2019A&A...623A...5D} {623, A5}

\bibitem[\protect\citeauthoryear{{Di Cesare}, {Graziani}, {Schneider},
  {Ginolfi}, {Venditti}, {Santini}  \& {Hunt}}{{Di Cesare}
  et~al.}{2023}]{dicesare23a}
{Di Cesare} C.,  {Graziani} L.,  {Schneider} R.,  {Ginolfi} M.,  {Venditti} A.,
   {Santini} P.,   {Hunt} L.~K.,  2023, \mn@doi [\mnras]
  {10.1093/mnras/stac3702}, \href
  {https://ui.adsabs.harvard.edu/abs/2023MNRAS.519.4632D} {519, 4632}

\bibitem[\protect\citeauthoryear{{Donnan} et~al.,}{{Donnan}
  et~al.}{2023a}]{donnan23a}
{Donnan} C.~T.,  et~al., 2023a, \mn@doi [\mnras] {10.1093/mnras/stac3472},
  \href {https://ui.adsabs.harvard.edu/abs/2023MNRAS.518.6011D} {518, 6011}

\bibitem[\protect\citeauthoryear{{Donnan}, {McLeod}, {McLure}, {Dunlop},
  {Carnall}, {Cullen}  \& {Magee}}{{Donnan} et~al.}{2023b}]{donnan23b}
{Donnan} C.~T.,  {McLeod} D.~J.,  {McLure} R.~J.,  {Dunlop} J.~S.,  {Carnall}
  A.~C.,  {Cullen} F.,   {Magee} D.,  2023b, \mn@doi [\mnras]
  {10.1093/mnras/stad471}, \href
  {https://ui.adsabs.harvard.edu/abs/2023MNRAS.520.4554D} {520, 4554}

\bibitem[\protect\citeauthoryear{{Dressler} et~al.,}{{Dressler}
  et~al.}{2023}]{dressler23a}
{Dressler} A.,  et~al., 2023, \mn@doi [arXiv/2306.02469]
  {10.48550/arXiv.2306.02469}, \href
  {https://ui.adsabs.harvard.edu/abs/2023arXiv230602469D} {p. arXiv:2306.02469}

\bibitem[\protect\citeauthoryear{{Dunlop}, {McLure}, {Robertson}, {Ellis},
  {Stark}, {Cirasuolo}  \& {de Ravel}}{{Dunlop} et~al.}{2012}]{dunlop12a}
{Dunlop} J.~S.,  {McLure} R.~J.,  {Robertson} B.~E.,  {Ellis} R.~S.,  {Stark}
  D.~P.,  {Cirasuolo} M.,   {de Ravel} L.,  2012, \mn@doi [\mnras]
  {10.1111/j.1365-2966.2011.20102.x}, \href
  {https://ui.adsabs.harvard.edu/abs/2012MNRAS.420..901D} {420, 901}

\bibitem[\protect\citeauthoryear{{Dwek}, {Galliano}  \& {Jones}}{{Dwek}
  et~al.}{2007}]{dwek07a}
{Dwek} E.,  {Galliano} F.,   {Jones} A.~P.,  2007, \mn@doi [\apj]
  {10.1086/518430}, \href
  {https://ui.adsabs.harvard.edu/abs/2007ApJ...662..927D} {662, 927}

\bibitem[\protect\citeauthoryear{{Endsley}, {Stark}, {Whitler}, {Topping},
  {Chen}, {Plat}, {Chisholm}  \& {Charlot}}{{Endsley}
  et~al.}{2022}]{endsley22a}
{Endsley} R.,  {Stark} D.~P.,  {Whitler} L.,  {Topping} M.~W.,  {Chen} Z.,
  {Plat} A.,  {Chisholm} J.,   {Charlot} S.,  2022, \mn@doi [arXiv/2208.14999]
  {10.48550/arXiv.2208.14999}, \href
  {https://ui.adsabs.harvard.edu/abs/2022arXiv220814999E} {p. arXiv:2208.14999}

\bibitem[\protect\citeauthoryear{{Endsley} et~al.,}{{Endsley}
  et~al.}{2023}]{endsley23a}
{Endsley} R.,  et~al., 2023, \mn@doi [arXiv e-prints]
  {10.48550/arXiv.2306.05295}, \href
  {https://ui.adsabs.harvard.edu/abs/2023arXiv230605295E} {p. arXiv:2306.05295}

\bibitem[\protect\citeauthoryear{{Esmerian} \& {Gnedin}}{{Esmerian} \&
  {Gnedin}}{2022}]{esmerian22a}
{Esmerian} C.~J.,  {Gnedin} N.~Y.,  2022, arXiv/2208.02277, \href
  {https://ui.adsabs.harvard.edu/abs/2022arXiv220802277E} {p. arXiv:2208.02277}

\bibitem[\protect\citeauthoryear{{Esmerian} \& {Gnedin}}{{Esmerian} \&
  {Gnedin}}{2023}]{esmerian23a}
{Esmerian} C.~J.,  {Gnedin} N.~Y.,  2023, \mn@doi [arXiv/2308.11723]
  {10.48550/arXiv.2308.11723}, \href
  {https://ui.adsabs.harvard.edu/abs/2023arXiv230811723E} {p. arXiv:2308.11723}

\bibitem[\protect\citeauthoryear{{Ferland} et~al.,}{{Ferland}
  et~al.}{2013}]{ferland13a}
{Ferland} G.~J.,  et~al., 2013, RMXAA, \href
  {http://adsabs.harvard.edu/abs/2013RMxAA..49..137F} {49, 137}

\bibitem[\protect\citeauthoryear{{Ferrara}}{{Ferrara}}{2024a}]{ferrara24c}
{Ferrara} A.,  2024a, \mn@doi [\aap] {10.1051/0004-6361/202348321}, \href
  {https://ui.adsabs.harvard.edu/abs/2024A&A...684A.207F} {684, A207}

\bibitem[\protect\citeauthoryear{{Ferrara}}{{Ferrara}}{2024b}]{ferrara24b}
{Ferrara} A.,  2024b, \mn@doi [\aap] {10.1051/0004-6361/202450944}, \href
  {https://ui.adsabs.harvard.edu/abs/2024A&A...689A.310F} {689, A310}

\bibitem[\protect\citeauthoryear{{Ferrara} et~al.,}{{Ferrara}
  et~al.}{2022}]{ferrara22a}
{Ferrara} A.,  et~al., 2022, \mn@doi [\mnras] {10.1093/mnras/stac460}, \href
  {https://ui.adsabs.harvard.edu/abs/2022MNRAS.512...58F} {512, 58}

\bibitem[\protect\citeauthoryear{{Ferrara}, {Pallottini}  \& {Dayal}}{{Ferrara}
  et~al.}{2023}]{ferrara23a}
{Ferrara} A.,  {Pallottini} A.,   {Dayal} P.,  2023, \mn@doi [\mnras]
  {10.1093/mnras/stad1095}, \href
  {https://ui.adsabs.harvard.edu/abs/2023MNRAS.522.3986F} {522, 3986}

\bibitem[\protect\citeauthoryear{{Ferrara}, {Carniani}, {di Mascia}, {Bouwens},
  {Oesch}  \& {Schouws}}{{Ferrara} et~al.}{2024}]{ferrara24a}
{Ferrara} A.,  {Carniani} S.,  {di Mascia} F.,  {Bouwens} R.,  {Oesch} P.,
  {Schouws} S.,  2024, \mn@doi [arXiv/2409.17223] {10.48550/arXiv.2409.17223},
  \href {https://ui.adsabs.harvard.edu/abs/2024arXiv240917223F} {p.
  arXiv:2409.17223}

\bibitem[\protect\citeauthoryear{{Ferrara}, {Carniani}, {di Mascia}, {Bouwens},
  {Oesch}  \& {Schouws}}{{Ferrara} et~al.}{2025a}]{ferrara25b}
{Ferrara} A.,  {Carniani} S.,  {di Mascia} F.,  {Bouwens} R.~J.,  {Oesch} P.,
  {Schouws} S.,  2025a, \mn@doi [\aap] {10.1051/0004-6361/202452368}, \href
  {https://ui.adsabs.harvard.edu/abs/2025A&A...694A.215F} {694, A215}

\bibitem[\protect\citeauthoryear{{Ferrara}, {Pallottini}  \&
  {Sommovigo}}{{Ferrara} et~al.}{2025b}]{ferrara25a}
{Ferrara} A.,  {Pallottini} A.,   {Sommovigo} L.,  2025b, \mn@doi [\aap]
  {10.1051/0004-6361/202452707}, \href
  {https://ui.adsabs.harvard.edu/abs/2025A&A...694A.286F} {694, A286}

\bibitem[\protect\citeauthoryear{{Finkelstein} et~al.,}{{Finkelstein}
  et~al.}{2012}]{finkelstein12a}
{Finkelstein} S.~L.,  et~al., 2012, \mn@doi [\apj]
  {10.1088/0004-637X/756/2/164}, \href
  {http://adsabs.harvard.edu/abs/2012ApJ...756..164F} {756, 164}

\bibitem[\protect\citeauthoryear{{Finkelstein} et~al.,}{{Finkelstein}
  et~al.}{2023}]{finkelstein23a}
{Finkelstein} S.~L.,  et~al., 2023, \mn@doi [\apjl] {10.3847/2041-8213/acade4},
  \href {https://ui.adsabs.harvard.edu/abs/2023ApJ...946L..13F} {946, L13}

\bibitem[\protect\citeauthoryear{{Fiore}, {Ferrara}, {Bischetti}, {Feruglio}
  \& {Travascio}}{{Fiore} et~al.}{2023}]{fiore23a}
{Fiore} F.,  {Ferrara} A.,  {Bischetti} M.,  {Feruglio} C.,   {Travascio} A.,
  2023, \mn@doi [\apjl] {10.3847/2041-8213/acb5f2}, \href
  {https://ui.adsabs.harvard.edu/abs/2023ApJ...943L..27F} {943, L27}

\bibitem[\protect\citeauthoryear{{Fudamoto} et~al.,}{{Fudamoto}
  et~al.}{2021}]{fudamoto21a}
{Fudamoto} Y.,  et~al., 2021, \mn@doi [\nat] {10.1038/s41586-021-03846-z},
  \href {https://ui.adsabs.harvard.edu/abs/2021Natur.597..489F} {597, 489}

\bibitem[\protect\citeauthoryear{{Garcia} et~al.,}{{Garcia}
  et~al.}{2024}]{garcia24a}
{Garcia} A.~M.,  et~al., 2024, \mn@doi [\mnras] {10.1093/mnras/stae1252}, \href
  {https://ui.adsabs.harvard.edu/abs/2024MNRAS.531.1398G} {531, 1398}

\bibitem[\protect\citeauthoryear{{Garcia} et~al.,}{{Garcia}
  et~al.}{2025}]{garcia25a}
{Garcia} A.~M.,  et~al., 2025, \mn@doi [\mnras] {10.1093/mnras/stae2587}, \href
  {https://ui.adsabs.harvard.edu/abs/2025MNRAS.536..119G} {536, 119}

\bibitem[\protect\citeauthoryear{{Garg}, {Narayanan}, {Sanders}, {Dav{\`e}},
  {Popping}, {Shapley}, {Stark}  \& {Trump}}{{Garg} et~al.}{2023}]{garg23a}
{Garg} P.,  {Narayanan} D.,  {Sanders} R.~L.,  {Dav{\`e}} R.,  {Popping} G.,
  {Shapley} A.~E.,  {Stark} D.~P.,   {Trump} J.~R.,  2023, \mn@doi
  [arXiv/2310.08622] {10.48550/arXiv.2310.08622}, \href
  {https://ui.adsabs.harvard.edu/abs/2023arXiv231008622G} {p. arXiv:2310.08622}

\bibitem[\protect\citeauthoryear{{Gilda}, {Lower}  \& {Narayanan}}{{Gilda}
  et~al.}{2021}]{gilda21a}
{Gilda} S.,  {Lower} S.,   {Narayanan} D.,  2021, \mn@doi [\apj]
  {10.3847/1538-4357/ac0058}, \href
  {https://ui.adsabs.harvard.edu/abs/2021ApJ...916...43G} {916, 43}

\bibitem[\protect\citeauthoryear{{Graziani}, {Schneider}, {Ginolfi}, {Hunt},
  {Maio}, {Glatzle}  \& {Ciardi}}{{Graziani} et~al.}{2020}]{graziani20a}
{Graziani} L.,  {Schneider} R.,  {Ginolfi} M.,  {Hunt} L.~K.,  {Maio} U.,
  {Glatzle} M.,   {Ciardi} B.,  2020, \mn@doi [\mnras] {10.1093/mnras/staa796},
  \href {https://ui.adsabs.harvard.edu/abs/2020MNRAS.494.1071G} {494, 1071}

\bibitem[\protect\citeauthoryear{{Guo} \& {White}}{{Guo} \&
  {White}}{2008}]{guo08a}
{Guo} Q.,  {White} S.~D.~M.,  2008, \mn@doi [\mnras]
  {10.1111/j.1365-2966.2007.12619.x}, \href
  {http://adsabs.harvard.edu/abs/2008MNRAS.384....2G} {384, 2}

\bibitem[\protect\citeauthoryear{{Gurvich} et~al.,}{{Gurvich}
  et~al.}{2023}]{gurvich23a}
{Gurvich} A.~B.,  et~al., 2023, \mn@doi [\mnras] {10.1093/mnras/stac3712},
  \href {https://ui.adsabs.harvard.edu/abs/2023MNRAS.519.2598G} {519, 2598}

\bibitem[\protect\citeauthoryear{{Hahn} \& {Abel}}{{Hahn} \&
  {Abel}}{2011}]{hahn11a}
{Hahn} O.,  {Abel} T.,  2011, \mn@doi [\mnras]
  {10.1111/j.1365-2966.2011.18820.x}, \href
  {http://adsabs.harvard.edu/abs/2011MNRAS.415.2101H} {415, 2101}

\bibitem[\protect\citeauthoryear{{Harikane} et~al.,}{{Harikane}
  et~al.}{2023}]{harikane23a}
{Harikane} Y.,  et~al., 2023, \mn@doi [\apjs] {10.3847/1538-4365/acaaa9}, \href
  {https://ui.adsabs.harvard.edu/abs/2023ApJS..265....5H} {265, 5}

\bibitem[\protect\citeauthoryear{{Harvey} et~al.,}{{Harvey}
  et~al.}{2025}]{harvey25a}
{Harvey} T.,  et~al., 2025, \mn@doi [arXiv e-prints]
  {10.48550/arXiv.2504.05244}, \href
  {https://ui.adsabs.harvard.edu/abs/2025arXiv250405244H} {p. arXiv:2504.05244}

\bibitem[\protect\citeauthoryear{{Hashimoto} et~al.,}{{Hashimoto}
  et~al.}{2019}]{hashimoto19a}
{Hashimoto} T.,  et~al., 2019, \mn@doi [\pasj] {10.1093/pasj/psz049}, \href
  {https://ui.adsabs.harvard.edu/abs/2019PASJ...71...71H} {71, 71}

\bibitem[\protect\citeauthoryear{{Heintz} et~al.,}{{Heintz}
  et~al.}{2023}]{heintz23a}
{Heintz} K.~E.,  et~al., 2023, \mn@doi [\aap] {10.1051/0004-6361/202347418},
  \href {https://ui.adsabs.harvard.edu/abs/2023A&A...679A..91H} {679, A91}

\bibitem[\protect\citeauthoryear{{Hirashita} \& {Yan}}{{Hirashita} \&
  {Yan}}{2009}]{hirashita09a}
{Hirashita} H.,  {Yan} H.,  2009, \mn@doi [\mnras]
  {10.1111/j.1365-2966.2009.14405.x}, \href
  {https://ui.adsabs.harvard.edu/abs/2009MNRAS.394.1061H} {394, 1061}

\bibitem[\protect\citeauthoryear{{Hopkins}}{{Hopkins}}{2014}]{hopkins14a}
{Hopkins} P.~F.,  2014, arXiv/1409.7395, \href
  {http://adsabs.harvard.edu/abs/2014arXiv1409.7395H} {}

\bibitem[\protect\citeauthoryear{{Hopkins} et~al.,}{{Hopkins}
  et~al.}{2018}]{hopkins18a}
{Hopkins} P.~F.,  et~al., 2018, \mn@doi [\mnras] {10.1093/mnras/sty1690}, \href
  {https://ui.adsabs.harvard.edu/abs/2018MNRAS.480..800H} {480, 800}

\bibitem[\protect\citeauthoryear{{Jaacks}, {Thompson}, {Finkelstein}  \&
  {Bromm}}{{Jaacks} et~al.}{2018}]{jaacks18a}
{Jaacks} J.,  {Thompson} R.,  {Finkelstein} S.~L.,   {Bromm} V.,  2018, \mn@doi
  [\mnras] {10.1093/mnras/sty062}, \href
  {https://ui.adsabs.harvard.edu/abs/2018MNRAS.475.4396J} {475, 4396}

\bibitem[\protect\citeauthoryear{{Jones}, {Tielens}  \& {Hollenbach}}{{Jones}
  et~al.}{1996}]{jones96a}
{Jones} A.~P.,  {Tielens} A.~G.~G.~M.,   {Hollenbach} D.~J.,  1996, \mn@doi
  [\apj] {10.1086/177823}, \href
  {https://ui.adsabs.harvard.edu/abs/1996ApJ...469..740J} {469, 740}

\bibitem[\protect\citeauthoryear{{Jones}, {Smith}, {Dav{\'e}}, {Narayanan}  \&
  {Li}}{{Jones} et~al.}{2024}]{jones24a}
{Jones} E.,  {Smith} B.,  {Dav{\'e}} R.,  {Narayanan} D.,   {Li} Q.,  2024,
  \mn@doi [\mnras] {10.1093/mnras/stae2445}, \href
  {https://ui.adsabs.harvard.edu/abs/2024MNRAS.535.1293J} {535, 1293}

\bibitem[\protect\citeauthoryear{{Katz}, {Weinberg}  \& {Hernquist}}{{Katz}
  et~al.}{1996}]{katz96a}
{Katz} N.,  {Weinberg} D.~H.,   {Hernquist} L.,  1996, \mn@doi [\apjs]
  {10.1086/192305}, \href {http://adsabs.harvard.edu/abs/1996ApJS..105...19K}
  {105, 19}

\bibitem[\protect\citeauthoryear{{Katz} et~al.,}{{Katz} et~al.}{2024}]{katz24a}
{Katz} H.,  et~al., 2024, arXiv/2408.03189, \href
  {https://ui.adsabs.harvard.edu/abs/2024arXiv240803189K} {p. arXiv:2408.03189}

\bibitem[\protect\citeauthoryear{{Kennicutt}}{{Kennicutt}}{1998}]{kennicutt98a}
{Kennicutt} Jr. R.~C.,  1998, \mn@doi [\araa] {10.1146/annurev.astro.36.1.189},
  \href {http://adsabs.harvard.edu/abs/1998ARA%26A..36..189K} {36, 189}

\bibitem[\protect\citeauthoryear{{Knudsen}, {Watson}, {Frayer}, {Christensen},
  {Gallazzi}, {Micha{\l}owski}, {Richard}  \& {Zavala}}{{Knudsen}
  et~al.}{2017}]{knudsen17a}
{Knudsen} K.~K.,  {Watson} D.,  {Frayer} D.,  {Christensen} L.,  {Gallazzi} A.,
   {Micha{\l}owski} M.~J.,  {Richard} J.,   {Zavala} J.,  2017, \mn@doi
  [\mnras] {10.1093/mnras/stw3066}, \href
  {https://ui.adsabs.harvard.edu/abs/2017MNRAS.466..138K} {466, 138}

\bibitem[\protect\citeauthoryear{{Krumholz} \& {Tan}}{{Krumholz} \&
  {Tan}}{2007}]{krumholz07b}
{Krumholz} M.~R.,  {Tan} J.~C.,  2007, \mn@doi [\apj] {10.1086/509101}, \href
  {http://adsabs.harvard.edu/abs/2007ApJ...654..304K} {654, 304}

\bibitem[\protect\citeauthoryear{{Krumholz}, {McKee}  \&
  {Tumlinson}}{{Krumholz} et~al.}{2008}]{krumholz08a}
{Krumholz} M.~R.,  {McKee} C.~F.,   {Tumlinson} J.,  2008, \mn@doi [\apj]
  {10.1086/592490}, \href {http://adsabs.harvard.edu/abs/2008ApJ...689..865K}
  {689, 865}

\bibitem[\protect\citeauthoryear{{Langeroodi} et~al.,}{{Langeroodi}
  et~al.}{2023}]{langeroodi23a}
{Langeroodi} D.,  et~al., 2023, \mn@doi [\apj] {10.3847/1538-4357/acdbc1},
  \href {https://ui.adsabs.harvard.edu/abs/2023ApJ...957...39L} {957, 39}

\bibitem[\protect\citeauthoryear{{Laporte} et~al.,}{{Laporte}
  et~al.}{2017}]{laporte17a}
{Laporte} N.,  et~al., 2017, \mn@doi [\apjl] {10.3847/2041-8213/aa62aa}, \href
  {https://ui.adsabs.harvard.edu/abs/2017ApJ...837L..21L} {837, L21}

\bibitem[\protect\citeauthoryear{{Lewis}, {Ocvirk}, {Dubois}, {Aubert},
  {Chardin}, {Gillet}  \& {Th{\'e}lie}}{{Lewis} et~al.}{2023}]{lewis23a}
{Lewis} J. S.~W.,  {Ocvirk} P.,  {Dubois} Y.,  {Aubert} D.,  {Chardin} J.,
  {Gillet} N.,   {Th{\'e}lie} {\'E}.,  2023, \mn@doi [\mnras]
  {10.1093/mnras/stad081}, \href
  {https://ui.adsabs.harvard.edu/abs/2023MNRAS.519.5987L} {519, 5987}

\bibitem[\protect\citeauthoryear{{Li}, {Narayanan}, {Dav{\`e}}  \&
  {Krumholz}}{{Li} et~al.}{2018}]{li18a}
{Li} Q.,  {Narayanan} D.,  {Dav{\`e}} R.,   {Krumholz} M.~R.,  2018, \mn@doi
  [\apj] {10.3847/1538-4357/aaec77}, \href
  {http://adsabs.harvard.edu/abs/2018ApJ...869...73L} {869, 73}

\bibitem[\protect\citeauthoryear{{Li}, {Narayanan}  \& {Dav{\'e}}}{{Li}
  et~al.}{2019}]{li19a}
{Li} Q.,  {Narayanan} D.,   {Dav{\'e}} R.,  2019, arXiv e-prints, \href
  {https://ui.adsabs.harvard.edu/abs/2019arXiv190609277L} {p. arXiv:1906.09277}

\bibitem[\protect\citeauthoryear{{Li}, {Narayanan}, {Torrey}, {Dav{\'e}}  \&
  {Vogelsberger}}{{Li} et~al.}{2021}]{li21a}
{Li} Q.,  {Narayanan} D.,  {Torrey} P.,  {Dav{\'e}} R.,   {Vogelsberger} M.,
  2021, \mn@doi [\mnras] {10.1093/mnras/stab2196}, \href
  {https://ui.adsabs.harvard.edu/abs/2021MNRAS.507..548L} {507, 548}

\bibitem[\protect\citeauthoryear{{Lower}, {Narayanan}, {Li}  \&
  {Dav{\'e}}}{{Lower} et~al.}{2023}]{lower23a}
{Lower} S.,  {Narayanan} D.,  {Li} Q.,   {Dav{\'e}} R.,  2023, \mn@doi [\apj]
  {10.3847/1538-4357/accf8c}, \href
  {https://ui.adsabs.harvard.edu/abs/2023ApJ...950...94L} {950, 94}

\bibitem[\protect\citeauthoryear{{Lower}, {Narayanan}, {Hu}  \&
  {Privon}}{{Lower} et~al.}{2024}]{lower24a}
{Lower} S.,  {Narayanan} D.,  {Hu} C.-Y.,   {Privon} G.~C.,  2024, \mn@doi
  [\apj] {10.3847/1538-4357/ad306c}, \href
  {https://ui.adsabs.harvard.edu/abs/2024ApJ...965..123L} {965, 123}

\bibitem[\protect\citeauthoryear{{Ma}, {Hopkins}, {Faucher-Gigu{\`e}re},
  {Zolman}, {Muratov}, {Kere{\v{s}}}  \& {Quataert}}{{Ma} et~al.}{2016}]{ma16a}
{Ma} X.,  {Hopkins} P.~F.,  {Faucher-Gigu{\`e}re} C.-A.,  {Zolman} N.,
  {Muratov} A.~L.,  {Kere{\v{s}}} D.,   {Quataert} E.,  2016, \mn@doi [\mnras]
  {10.1093/mnras/stv2659}, \href
  {https://ui.adsabs.harvard.edu/abs/2016MNRAS.456.2140M} {456, 2140}

\bibitem[\protect\citeauthoryear{{Magdis} et~al.,}{{Magdis}
  et~al.}{2012}]{magdis12a}
{Magdis} G.~E.,  et~al., 2012, \mn@doi [\apj] {10.1088/0004-637X/760/1/6},
  \href {https://ui.adsabs.harvard.edu/abs/2012ApJ...760....6M} {760, 6}

\bibitem[\protect\citeauthoryear{{Marinacci}, {Sales}, {Vogelsberger}, {Torrey}
   \& {Springel}}{{Marinacci} et~al.}{2019}]{marinacci19a}
{Marinacci} F.,  {Sales} L.~V.,  {Vogelsberger} M.,  {Torrey} P.,   {Springel}
  V.,  2019, \mn@doi [\mnras] {10.1093/mnras/stz2391}, \href
  {https://ui.adsabs.harvard.edu/abs/2019MNRAS.489.4233M} {489, 4233}

\bibitem[\protect\citeauthoryear{{Markov}, {Gallerani}, {Ferrara},
  {Pallottini}, {Parlanti}, {Di Mascia}, {Sommovigo}  \& {Kohandel}}{{Markov}
  et~al.}{2024}]{markov24a}
{Markov} V.,  {Gallerani} S.,  {Ferrara} A.,  {Pallottini} A.,  {Parlanti} E.,
  {Di Mascia} F.,  {Sommovigo} L.,   {Kohandel} M.,  2024, \mn@doi
  [arXiv/2402.05996] {10.48550/arXiv.2402.05996}, \href
  {https://ui.adsabs.harvard.edu/abs/2024arXiv240205996M} {p. arXiv:2402.05996}

\bibitem[\protect\citeauthoryear{{Marrone} et~al.,}{{Marrone}
  et~al.}{2018}]{marrone18a}
{Marrone} D.~P.,  et~al., 2018, \mn@doi [\nat] {10.1038/nature24629}, \href
  {https://ui.adsabs.harvard.edu/abs/2018Natur.553...51M} {553, 51}

\bibitem[\protect\citeauthoryear{{Mathis}, {Rumpl}  \& {Nordsieck}}{{Mathis}
  et~al.}{1977}]{mathis77a}
{Mathis} J.~S.,  {Rumpl} W.,   {Nordsieck} K.~H.,  1977, \mn@doi [\apj]
  {10.1086/155591}, \href
  {https://ui.adsabs.harvard.edu/abs/1977ApJ...217..425M} {217, 425}

\bibitem[\protect\citeauthoryear{{Matsuura} et~al.,}{{Matsuura}
  et~al.}{2011}]{matsuura11a}
{Matsuura} M.,  et~al., 2011, \mn@doi [Science] {10.1126/science.1205983},
  \href {https://ui.adsabs.harvard.edu/abs/2011Sci...333.1258M} {333, 1258}

\bibitem[\protect\citeauthoryear{{McClymont} et~al.,}{{McClymont}
  et~al.}{2025}]{mcclymont25a}
{McClymont} W.,  et~al., 2025, \mn@doi [arXiv/2507.08787]
  {10.48550/arXiv.2507.08787}, \href
  {https://ui.adsabs.harvard.edu/abs/2025arXiv250708787M} {p. arXiv:2507.08787}

\bibitem[\protect\citeauthoryear{{McKinney}, {Cooper}, {Casey}, {Mu{\~n}oz},
  {Akins}, {Lambrides}  \& {Long}}{{McKinney} et~al.}{2025}]{mckinney25a}
{McKinney} J.,  {Cooper} O.~R.,  {Casey} C.~M.,  {Mu{\~n}oz} J.~B.,  {Akins}
  H.,  {Lambrides} E.,   {Long} A.~S.,  2025, \mn@doi [\apjl]
  {10.3847/2041-8213/add15d}, \href
  {https://ui.adsabs.harvard.edu/abs/2025ApJ...985L..21M} {985, L21}

\bibitem[\protect\citeauthoryear{{McKinnon}, {Torrey}  \&
  {Vogelsberger}}{{McKinnon} et~al.}{2016}]{mckinnon16a}
{McKinnon} R.,  {Torrey} P.,   {Vogelsberger} M.,  2016, \mn@doi [\mnras]
  {10.1093/mnras/stw253}, \href
  {http://adsabs.harvard.edu/abs/2016MNRAS.457.3775M} {457, 3775}

\bibitem[\protect\citeauthoryear{{McKinnon}, {Torrey}, {Vogelsberger},
  {Hayward}  \& {Marinacci}}{{McKinnon} et~al.}{2017}]{mckinnon17a}
{McKinnon} R.,  {Torrey} P.,  {Vogelsberger} M.,  {Hayward} C.~C.,
  {Marinacci} F.,  2017, \mn@doi [\mnras] {10.1093/mnras/stx467}, \href
  {https://ui.adsabs.harvard.edu/abs/2017MNRAS.468.1505M} {468, 1505}

\bibitem[\protect\citeauthoryear{{McKinnon}, {Vogelsberger}, {Torrey},
  {Marinacci}  \& {Kannan}}{{McKinnon} et~al.}{2018}]{mckinnon18a}
{McKinnon} R.,  {Vogelsberger} M.,  {Torrey} P.,  {Marinacci} F.,   {Kannan}
  R.,  2018, \mn@doi [\mnras] {10.1093/mnras/sty1248}, \href
  {https://ui.adsabs.harvard.edu/abs/2018MNRAS.478.2851M} {478, 2851}

\bibitem[\protect\citeauthoryear{{Menon}, {Balu}  \& {Power}}{{Menon}
  et~al.}{2025}]{menon25a}
{Menon} A.,  {Balu} S.,   {Power} C.,  2025, \mn@doi [arXiv/2508.86363]
  {10.48550/arXiv.2508.08363}, \href
  {https://ui.adsabs.harvard.edu/abs/2025arXiv250808363M} {p. arXiv:2508.08363}

\bibitem[\protect\citeauthoryear{{Meurer}, {Heckman}  \& {Calzetti}}{{Meurer}
  et~al.}{1999}]{meurer99a}
{Meurer} G.~R.,  {Heckman} T.~M.,   {Calzetti} D.,  1999, \mn@doi [\apj]
  {10.1086/307523}, \href {http://adsabs.harvard.edu/abs/1999ApJ...521...64M}
  {521, 64}

\bibitem[\protect\citeauthoryear{{Micha{\l}owski}}{{Micha{\l}owski}}{2015}]{michalowski15a}
{Micha{\l}owski} M.~J.,  2015, \mn@doi [\aap] {10.1051/0004-6361/201525644},
  \href {https://ui.adsabs.harvard.edu/abs/2015A&A...577A..80M} {577, A80}

\bibitem[\protect\citeauthoryear{{Morales} et~al.,}{{Morales}
  et~al.}{2024}]{morales24a}
{Morales} A.~M.,  et~al., 2024, \mn@doi [\apjl] {10.3847/2041-8213/ad2de4},
  \href {https://ui.adsabs.harvard.edu/abs/2024ApJ...964L..24M} {964, L24}

\bibitem[\protect\citeauthoryear{{Mosleh}, {Riahi-Zamin}  \&
  {Tacchella}}{{Mosleh} et~al.}{2025}]{mosleh25a}
{Mosleh} M.,  {Riahi-Zamin} M.,   {Tacchella} S.,  2025, \mn@doi [\apj]
  {10.3847/1538-4357/adc12e}, \href
  {https://ui.adsabs.harvard.edu/abs/2025ApJ...983..181M} {983, 181}

\bibitem[\protect\citeauthoryear{{Muratov} et~al.,}{{Muratov}
  et~al.}{2017}]{muratov17a}
{Muratov} A.~L.,  et~al., 2017, \mn@doi [\mnras] {10.1093/mnras/stx667}, \href
  {https://ui.adsabs.harvard.edu/abs/2017MNRAS.468.4170M} {468, 4170}

\bibitem[\protect\citeauthoryear{{Nakajima}, {Ouchi}, {Isobe}, {Harikane},
  {Zhang}, {Ono}, {Umeda}  \& {Oguri}}{{Nakajima} et~al.}{2023}]{nakajima23a}
{Nakajima} K.,  {Ouchi} M.,  {Isobe} Y.,  {Harikane} Y.,  {Zhang} Y.,  {Ono}
  Y.,  {Umeda} H.,   {Oguri} M.,  2023, \mn@doi [\apjs]
  {10.3847/1538-4365/acd556}, \href
  {https://ui.adsabs.harvard.edu/abs/2023ApJS..269...33N} {269, 33}

\bibitem[\protect\citeauthoryear{{Narayanan}, {Dav{\'e}}, {Johnson},
  {Thompson}, {Conroy}  \& {Geach}}{{Narayanan} et~al.}{2018}]{narayanan18a}
{Narayanan} D.,  {Dav{\'e}} R.,  {Johnson} B.~D.,  {Thompson} R.,  {Conroy} C.,
    {Geach} J.,  2018, \mn@doi [\mnras] {10.1093/mnras/stx2860}, \href
  {http://adsabs.harvard.edu/abs/2018MNRAS.474.1718N} {474, 1718}

\bibitem[\protect\citeauthoryear{{Narayanan} et~al.,}{{Narayanan}
  et~al.}{2021}]{narayanan21a}
{Narayanan} D.,  et~al., 2021, \mn@doi [\apjs] {10.3847/1538-4365/abc487},
  \href {https://ui.adsabs.harvard.edu/abs/2021ApJS..252...12N} {252, 12}

\bibitem[\protect\citeauthoryear{{Narayanan} et~al.,}{{Narayanan}
  et~al.}{2023}]{narayanan23a}
{Narayanan} D.,  et~al., 2023, \mn@doi [\apj] {10.3847/1538-4357/accf8d}, \href
  {https://ui.adsabs.harvard.edu/abs/2023ApJ...951..100N} {951, 100}

\bibitem[\protect\citeauthoryear{{Narayanan} et~al.,}{{Narayanan}
  et~al.}{2024}]{narayanan24a}
{Narayanan} D.,  et~al., 2024, \mn@doi [\apj] {10.3847/1538-4357/ad0966}, \href
  {https://ui.adsabs.harvard.edu/abs/2024ApJ...961...73N} {961, 73}

\bibitem[\protect\citeauthoryear{{Narayanan} et~al.,}{{Narayanan}
  et~al.}{2025}]{narayanan25a}
{Narayanan} D.,  et~al., 2025, \mn@doi [\apj] {10.3847/1538-4357/adb41c}, \href
  {https://ui.adsabs.harvard.edu/abs/2025ApJ...982....7N} {982, 7}

\bibitem[\protect\citeauthoryear{{Ni} et~al.,}{{Ni} et~al.}{2023}]{ni23a}
{Ni} Y.,  et~al., 2023, \mn@doi [\apj] {10.3847/1538-4357/ad022a}, \href
  {https://ui.adsabs.harvard.edu/abs/2023ApJ...959..136N} {959, 136}

\bibitem[\protect\citeauthoryear{{Nomoto}, {Tominaga}, {Umeda}, {Kobayashi}  \&
  {Maeda}}{{Nomoto} et~al.}{2006}]{nomoto06a}
{Nomoto} K.,  {Tominaga} N.,  {Umeda} H.,  {Kobayashi} C.,   {Maeda} K.,  2006,
  \mn@doi [Nuclear Physics A] {10.1016/j.nuclphysa.2006.05.008}, \href
  {http://adsabs.harvard.edu/abs/2006NuPhA.777..424N} {777, 424}

\bibitem[\protect\citeauthoryear{{Nozawa}, {Kozasa}  \& {Habe}}{{Nozawa}
  et~al.}{2006}]{nozawa06a}
{Nozawa} T.,  {Kozasa} T.,   {Habe} A.,  2006, \mn@doi [\apj] {10.1086/505639},
  \href {https://ui.adsabs.harvard.edu/abs/2006ApJ...648..435N} {648, 435}

\bibitem[\protect\citeauthoryear{{Nozawa}, {Kozasa}, {Habe}, {Dwek}, {Umeda},
  {Tominaga}, {Maeda}  \& {Nomoto}}{{Nozawa} et~al.}{2007}]{nozawa07a}
{Nozawa} T.,  {Kozasa} T.,  {Habe} A.,  {Dwek} E.,  {Umeda} H.,  {Tominaga} N.,
   {Maeda} K.,   {Nomoto} K.,  2007, \mn@doi [\apj] {10.1086/520621}, \href
  {https://ui.adsabs.harvard.edu/abs/2007ApJ...666..955N} {666, 955}

\bibitem[\protect\citeauthoryear{{Nozawa}, {Kozasa}, {Tominaga}, {Maeda},
  {Umeda}, {Nomoto}  \& {Krause}}{{Nozawa} et~al.}{2010}]{nozawa10a}
{Nozawa} T.,  {Kozasa} T.,  {Tominaga} N.,  {Maeda} K.,  {Umeda} H.,  {Nomoto}
  K.,   {Krause} O.,  2010, \mn@doi [\apj] {10.1088/0004-637X/713/1/356}, \href
  {https://ui.adsabs.harvard.edu/abs/2010ApJ...713..356N} {713, 356}

\bibitem[\protect\citeauthoryear{{Ormerod} et~al.,}{{Ormerod}
  et~al.}{2025}]{ormerod25a}
{Ormerod} K.,  et~al., 2025, \mn@doi [arXiv/2502/21119]
  {10.48550/arXiv.2502.21119}, \href
  {https://ui.adsabs.harvard.edu/abs/2025arXiv250221119O} {p. arXiv:2502.21119}

\bibitem[\protect\citeauthoryear{{Parente}, {Ragone-Figueroa}, {Granato},
  {Borgani}, {Murante}, {Valentini}, {Bressan}  \& {Lapi}}{{Parente}
  et~al.}{2022}]{parente22a}
{Parente} M.,  {Ragone-Figueroa} C.,  {Granato} G.~L.,  {Borgani} S.,
  {Murante} G.,  {Valentini} M.,  {Bressan} A.,   {Lapi} A.,  2022, \mn@doi
  [\mnras] {10.1093/mnras/stac1913}, \href
  {https://ui.adsabs.harvard.edu/abs/2022MNRAS.515.2053P} {515, 2053}

\bibitem[\protect\citeauthoryear{{Parente}, {Ragone-Figueroa}, {Granato}  \&
  {Lapi}}{{Parente} et~al.}{2023}]{parente23a}
{Parente} M.,  {Ragone-Figueroa} C.,  {Granato} G.~L.,   {Lapi} A.,  2023,
  \mn@doi [\mnras] {10.1093/mnras/stad907}, \href
  {https://ui.adsabs.harvard.edu/abs/2023MNRAS.521.6105P} {521, 6105}

\bibitem[\protect\citeauthoryear{{Popping} \& {P{\'e}roux}}{{Popping} \&
  {P{\'e}roux}}{2022}]{popping22a}
{Popping} G.,  {P{\'e}roux} C.,  2022, \mn@doi [\mnras]
  {10.1093/mnras/stac695}, \href
  {https://ui.adsabs.harvard.edu/abs/2022MNRAS.513.1531P} {513, 1531}

\bibitem[\protect\citeauthoryear{{Popping}, {Puglisi}  \& {Norman}}{{Popping}
  et~al.}{2017}]{popping17a}
{Popping} G.,  {Puglisi} A.,   {Norman} C.~A.,  2017, \mn@doi [\mnras]
  {10.1093/mnras/stx2202}, \href
  {http://adsabs.harvard.edu/abs/2017MNRAS.472.2315P} {472, 2315}

\bibitem[\protect\citeauthoryear{{Popping} et~al.,}{{Popping}
  et~al.}{2023}]{popping23a}
{Popping} G.,  et~al., 2023, \mn@doi [\aap] {10.1051/0004-6361/202243817},
  \href {https://ui.adsabs.harvard.edu/abs/2023A&A...670A.138P} {670, A138}

\bibitem[\protect\citeauthoryear{{Rahmati}, {Pawlik}, {Rai{\v{c}}evi{\'c}}  \&
  {Schaye}}{{Rahmati} et~al.}{2013}]{rahmati13a}
{Rahmati} A.,  {Pawlik} A.~H.,  {Rai{\v{c}}evi{\'c}} M.,   {Schaye} J.,  2013,
  \mn@doi [\mnras] {10.1093/mnras/stt066}, \href
  {https://ui.adsabs.harvard.edu/abs/2013MNRAS.430.2427R} {430, 2427}

\bibitem[\protect\citeauthoryear{{Raiter}, {Schaerer}  \& {Fosbury}}{{Raiter}
  et~al.}{2010}]{raiter10a}
{Raiter} A.,  {Schaerer} D.,   {Fosbury} R.~A.~E.,  2010, \mn@doi [\aap]
  {10.1051/0004-6361/201015236}, \href
  {https://ui.adsabs.harvard.edu/abs/2010A&A...523A..64R} {523, A64}

\bibitem[\protect\citeauthoryear{{Reddy} et~al.,}{{Reddy}
  et~al.}{2018}]{reddy18a}
{Reddy} N.~A.,  et~al., 2018, \mn@doi [\apj] {10.3847/1538-4357/aaa3e7}, \href
  {https://ui.adsabs.harvard.edu/abs/2018ApJ...853...56R} {853, 56}

\bibitem[\protect\citeauthoryear{{R{\'e}my-Ruyer} et~al.,}{{R{\'e}my-Ruyer}
  et~al.}{2014}]{remyruyer14a}
{R{\'e}my-Ruyer} A.,  et~al., 2014, \mn@doi [\aap]
  {10.1051/0004-6361/201322803}, \href
  {https://ui.adsabs.harvard.edu/abs/2014A&A...563A..31R} {563, A31}

\bibitem[\protect\citeauthoryear{{Roberts-Borsani} et~al.,}{{Roberts-Borsani}
  et~al.}{2024}]{roberts-borsani24a}
{Roberts-Borsani} G.,  et~al., 2024, \mn@doi [arXiv/2407.17551]
  {10.48550/arXiv.2407.17551}, \href
  {https://ui.adsabs.harvard.edu/abs/2024arXiv240717551R} {p. arXiv:2407.17551}

\bibitem[\protect\citeauthoryear{{Robertson} et~al.,}{{Robertson}
  et~al.}{2023}]{robertson23a}
{Robertson} B.~E.,  et~al., 2023, \mn@doi [Nature Astronomy]
  {10.1038/s41550-023-01921-1}, \href
  {https://ui.adsabs.harvard.edu/abs/2023NatAs...7..611R} {7, 611}

\bibitem[\protect\citeauthoryear{{Robitaille}}{{Robitaille}}{2011}]{robitaille11a}
{Robitaille} T.~P.,  2011, \mn@doi [\aap] {10.1051/0004-6361/201117150}, \href
  {http://adsabs.harvard.edu/abs/2011A%26A...536A..79R} {536, A79}

\bibitem[\protect\citeauthoryear{{Salim} \& {Narayanan}}{{Salim} \&
  {Narayanan}}{2020}]{salim20a}
{Salim} S.,  {Narayanan} D.,  2020, \mn@doi [\araa]
  {10.1146/annurev-astro-032620-021933}, \href
  {https://ui.adsabs.harvard.edu/abs/2020ARA&A..58..529S} {58, 529}

\bibitem[\protect\citeauthoryear{{Sanders} et~al.,}{{Sanders}
  et~al.}{2025}]{sanders25a}
{Sanders} R.~L.,  et~al., 2025, \mn@doi [arXiv/2508.10099]
  {10.48550/arXiv.2508.10099}, \href
  {https://ui.adsabs.harvard.edu/abs/2025arXiv250810099S} {p. arXiv:2508.10099}

\bibitem[\protect\citeauthoryear{{Schneider} \& {Maiolino}}{{Schneider} \&
  {Maiolino}}{2024}]{schneider24a}
{Schneider} R.,  {Maiolino} R.,  2024, \mn@doi [\aapr]
  {10.1007/s00159-024-00151-2}, \href
  {https://ui.adsabs.harvard.edu/abs/2024A&ARv..32....2S} {32, 2}

\bibitem[\protect\citeauthoryear{{Schneider}, {Valiante}, {Ventura},
  {dell'Agli}, {Di Criscienzo}, {Hirashita}  \& {Kemper}}{{Schneider}
  et~al.}{2014}]{schneider14a}
{Schneider} R.,  {Valiante} R.,  {Ventura} P.,  {dell'Agli} F.,  {Di
  Criscienzo} M.,  {Hirashita} H.,   {Kemper} F.,  2014, \mn@doi [\mnras]
  {10.1093/mnras/stu861}, \href
  {https://ui.adsabs.harvard.edu/abs/2014MNRAS.442.1440S} {442, 1440}

\bibitem[\protect\citeauthoryear{{Shapley}}{{Shapley}}{1953}]{shapley53a}
{Shapley} L.,  1953, PNAS, \href
  {http://adsabs.harvard.edu/abs/2004ApJ...612..108S} {39, 1095}

\bibitem[\protect\citeauthoryear{{Shchekinov} \& {Nath}}{{Shchekinov} \&
  {Nath}}{2025}]{shchekinov25a}
{Shchekinov} Y.~A.,  {Nath} B.~B.,  2025, \mn@doi [arXiv/2506.05591]
  {10.48550/arXiv.2506.05591}, \href
  {https://ui.adsabs.harvard.edu/abs/2025arXiv250605591S} {p. arXiv:2506.05591}

\bibitem[\protect\citeauthoryear{{Shen}, {Vogelsberger}, {Boylan-Kolchin},
  {Tacchella}  \& {Kannan}}{{Shen} et~al.}{2023}]{shen23a}
{Shen} X.,  {Vogelsberger} M.,  {Boylan-Kolchin} M.,  {Tacchella} S.,
  {Kannan} R.,  2023, \mn@doi [arXiv/2305.05679] {10.48550/arXiv.2305.05679},
  \href {https://ui.adsabs.harvard.edu/abs/2023arXiv230505679S} {p.
  arXiv:2305.05679}

\bibitem[\protect\citeauthoryear{{Shen} et~al.,}{{Shen} et~al.}{2025}]{shen25a}
{Shen} X.,  et~al., 2025, \mn@doi [arXiv e-prints] {10.48550/arXiv.2503.01949},
  \href {https://ui.adsabs.harvard.edu/abs/2025arXiv250301949S} {p.
  arXiv:2503.01949}

\bibitem[\protect\citeauthoryear{{Sommovigo} \& {Algera}}{{Sommovigo} \&
  {Algera}}{2025}]{sommovigo25a}
{Sommovigo} L.,  {Algera} H.,  2025, arXiv/2505.20105, \href
  {https://ui.adsabs.harvard.edu/abs/2025arXiv250520105S} {p. arXiv:2505.20105}

\bibitem[\protect\citeauthoryear{{Sommovigo} et~al.,}{{Sommovigo}
  et~al.}{2022}]{sommovigo22a}
{Sommovigo} L.,  et~al., 2022, \mn@doi [\mnras] {10.1093/mnras/stac302}, \href
  {https://ui.adsabs.harvard.edu/abs/2022MNRAS.513.3122S} {513, 3122}

\bibitem[\protect\citeauthoryear{{Sparre}, {Hayward}, {Feldmann},
  {Faucher-Gigu{\`e}re}, {Muratov}, {Kere{\v{s}}}  \& {Hopkins}}{{Sparre}
  et~al.}{2017}]{sparre17a}
{Sparre} M.,  {Hayward} C.~C.,  {Feldmann} R.,  {Faucher-Gigu{\`e}re} C.-A.,
  {Muratov} A.~L.,  {Kere{\v{s}}} D.,   {Hopkins} P.~F.,  2017, \mn@doi
  [\mnras] {10.1093/mnras/stw3011}, \href
  {https://ui.adsabs.harvard.edu/abs/2017MNRAS.466...88S} {466, 88}

\bibitem[\protect\citeauthoryear{{Springel}}{{Springel}}{2010}]{springel10a}
{Springel} V.,  2010, \mn@doi [\mnras] {10.1111/j.1365-2966.2009.15715.x},
  \href {http://adsabs.harvard.edu/abs/2010MNRAS.401..791S} {401, 791}

\bibitem[\protect\citeauthoryear{{Stanway}, {Eldridge}  \& {Becker}}{{Stanway}
  et~al.}{2016}]{stanway16a}
{Stanway} E.~R.,  {Eldridge} J.~J.,   {Becker} G.~D.,  2016, \mn@doi [\mnras]
  {10.1093/mnras/stv2661}, \href
  {http://adsabs.harvard.edu/abs/2016MNRAS.456..485S} {456, 485}

\bibitem[\protect\citeauthoryear{{Sun}, {Faucher-Gigu{\`e}re}, {Hayward},
  {Shen}, {Wetzel}  \& {Cochrane}}{{Sun} et~al.}{2023}]{sun23a}
{Sun} G.,  {Faucher-Gigu{\`e}re} C.-A.,  {Hayward} C.~C.,  {Shen} X.,  {Wetzel}
  A.,   {Cochrane} R.~K.,  2023, \mn@doi [arXiv/2307.15305]
  {10.48550/arXiv.2307.15305}, \href
  {https://ui.adsabs.harvard.edu/abs/2023arXiv230715305S} {p. arXiv:2307.15305}

\bibitem[\protect\citeauthoryear{{Tacchella} et~al.,}{{Tacchella}
  et~al.}{2022}]{tacchella22a}
{Tacchella} S.,  et~al., 2022, \mn@doi [\apj] {10.3847/1538-4357/ac4cad}, \href
  {https://ui.adsabs.harvard.edu/abs/2022ApJ...927..170T} {927, 170}

\bibitem[\protect\citeauthoryear{{Thompson}, {Nagamine}, {Jaacks}  \&
  {Choi}}{{Thompson} et~al.}{2014}]{thompson14a}
{Thompson} R.,  {Nagamine} K.,  {Jaacks} J.,   {Choi} J.-H.,  2014, \mn@doi
  [\apj] {10.1088/0004-637X/780/2/145}, \href
  {http://adsabs.harvard.edu/abs/2014ApJ...780..145T} {780, 145}

\bibitem[\protect\citeauthoryear{{Todini} \& {Ferrara}}{{Todini} \&
  {Ferrara}}{2001}]{todini01a}
{Todini} P.,  {Ferrara} A.,  2001, \mn@doi [\mnras]
  {10.1046/j.1365-8711.2001.04486.x}, \href
  {https://ui.adsabs.harvard.edu/abs/2001MNRAS.325..726T} {325, 726}

\bibitem[\protect\citeauthoryear{{Topping} et~al.,}{{Topping}
  et~al.}{2022}]{topping22a}
{Topping} M.~W.,  et~al., 2022, \mn@doi [\mnras] {10.1093/mnras/stac2291},
  \href {https://ui.adsabs.harvard.edu/abs/2022MNRAS.516..975T} {516, 975}

\bibitem[\protect\citeauthoryear{{Topping} et~al.,}{{Topping}
  et~al.}{2024}]{topping24a}
{Topping} M.~W.,  et~al., 2024, \mn@doi [\mnras] {10.1093/mnras/stae800}, \href
  {https://ui.adsabs.harvard.edu/abs/2024MNRAS.529.4087T} {529, 4087}

\bibitem[\protect\citeauthoryear{{Torrey}, {Vogelsberger}, {Genel}, {Sijacki},
  {Springel}  \& {Hernquist}}{{Torrey} et~al.}{2014}]{torrey14a}
{Torrey} P.,  {Vogelsberger} M.,  {Genel} S.,  {Sijacki} D.,  {Springel} V.,
  {Hernquist} L.,  2014, \mn@doi [\mnras] {10.1093/mnras/stt2295}, \href
  {https://ui.adsabs.harvard.edu/abs/2014MNRAS.438.1985T} {438, 1985}

\bibitem[\protect\citeauthoryear{{Trayford} et~al.,}{{Trayford}
  et~al.}{2025}]{trayford25a}
{Trayford} J.~W.,  et~al., 2025, \mn@doi [arXiv/2505.13056]
  {10.48550/arXiv.2505.13056}, \href
  {https://ui.adsabs.harvard.edu/abs/2025arXiv250513056T} {p. arXiv:2505.13056}

\bibitem[\protect\citeauthoryear{{Triani}, {Sinha}, {Croton}, {Pacifici}  \&
  {Dwek}}{{Triani} et~al.}{2020}]{triani20a}
{Triani} D.~P.,  {Sinha} M.,  {Croton} D.~J.,  {Pacifici} C.,   {Dwek} E.,
  2020, \mn@doi [\mnras] {10.1093/mnras/staa446}, \href
  {https://ui.adsabs.harvard.edu/abs/2020MNRAS.493.2490T} {493, 2490}

\bibitem[\protect\citeauthoryear{{Tsai} \& {Mathews}}{{Tsai} \&
  {Mathews}}{1995}]{tsai95a}
{Tsai} J.~C.,  {Mathews} W.~G.,  1995, \mn@doi [\apj] {10.1086/175943}, \href
  {https://ui.adsabs.harvard.edu/abs/1995ApJ...448...84T} {448, 84}

\bibitem[\protect\citeauthoryear{{Turk}, {Smith}, {Oishi}, {Skory}, {Skillman},
  {Abel}  \& {Norman}}{{Turk} et~al.}{2011}]{turk11a}
{Turk} M.~J.,  {Smith} B.~D.,  {Oishi} J.~S.,  {Skory} S.,  {Skillman} S.~W.,
  {Abel} T.,   {Norman} M.~L.,  2011, \mn@doi [\apjs]
  {10.1088/0067-0049/192/1/9}, \href
  {http://adsabs.harvard.edu/abs/2011ApJS..192....9T} {192, 9}

\bibitem[\protect\citeauthoryear{{Vijayan}, {Clay}, {Thomas}, {Yates},
  {Wilkins}  \& {Henriques}}{{Vijayan} et~al.}{2019}]{vijayan19a}
{Vijayan} A.~P.,  {Clay} S.~J.,  {Thomas} P.~A.,  {Yates} R.~M.,  {Wilkins}
  S.~M.,   {Henriques} B.~M.,  2019, \mn@doi [\mnras] {10.1093/mnras/stz1948},
  \href {https://ui.adsabs.harvard.edu/abs/2019MNRAS.489.4072V} {489, 4072}

\bibitem[\protect\citeauthoryear{{Vijayan} et~al.,}{{Vijayan}
  et~al.}{2022}]{vijayan22a}
{Vijayan} A.~P.,  et~al., 2022, \mn@doi [\mnras] {10.1093/mnras/stac338}, \href
  {https://ui.adsabs.harvard.edu/abs/2022MNRAS.511.4999V} {511, 4999}

\bibitem[\protect\citeauthoryear{{Vijayan} et~al.,}{{Vijayan}
  et~al.}{2025}]{vijayan25a}
{Vijayan} A.~P.,  et~al., 2025, arXiv2507.20190, \href
  {https://ui.adsabs.harvard.edu/abs/2025arXiv250720190V} {p. arXiv:2507.20190}

\bibitem[\protect\citeauthoryear{{Vogelsberger}, {Genel}, {Sijacki}, {Torrey},
  {Springel}  \& {Hernquist}}{{Vogelsberger} et~al.}{2013}]{vogelsberger13a}
{Vogelsberger} M.,  {Genel} S.,  {Sijacki} D.,  {Torrey} P.,  {Springel} V.,
  {Hernquist} L.,  2013, \mn@doi [\mnras] {10.1093/mnras/stt1789}, \href
  {https://ui.adsabs.harvard.edu/abs/2013MNRAS.436.3031V} {436, 3031}

\bibitem[\protect\citeauthoryear{{Wang} et~al.,}{{Wang} et~al.}{2025}]{wang25a}
{Wang} B.,  et~al., 2025, \mn@doi [arXiv/2504.15255]
  {10.48550/arXiv.2504.15255}, \href
  {https://ui.adsabs.harvard.edu/abs/2025arXiv250415255W} {p. arXiv:2504.15255}

\bibitem[\protect\citeauthoryear{{Watson}, {Christensen}, {Knudsen}, {Richard},
  {Gallazzi}  \& {Micha{\l}owski}}{{Watson} et~al.}{2015}]{watson15a}
{Watson} D.,  {Christensen} L.,  {Knudsen} K.~K.,  {Richard} J.,  {Gallazzi}
  A.,   {Micha{\l}owski} M.~J.,  2015, \mn@doi [\nat] {10.1038/nature14164},
  \href {http://adsabs.harvard.edu/abs/2015Natur.519..327W} {519, 327}

\bibitem[\protect\citeauthoryear{{Weinberger}, {Springel}  \&
  {Pakmor}}{{Weinberger} et~al.}{2020}]{weinberger20a}
{Weinberger} R.,  {Springel} V.,   {Pakmor} R.,  2020, \mn@doi [\apjs]
  {10.3847/1538-4365/ab908c}, \href
  {https://ui.adsabs.harvard.edu/abs/2020ApJS..248...32W} {248, 32}

\bibitem[\protect\citeauthoryear{{Whitler}, {Endsley}, {Stark}, {Topping},
  {Chen}  \& {Charlot}}{{Whitler} et~al.}{2023}]{whitler23a}
{Whitler} L.,  {Endsley} R.,  {Stark} D.~P.,  {Topping} M.,  {Chen} Z.,
  {Charlot} S.,  2023, \mn@doi [\mnras] {10.1093/mnras/stac3535}, \href
  {https://ui.adsabs.harvard.edu/abs/2023MNRAS.519..157W} {519, 157}

\bibitem[\protect\citeauthoryear{{Wilkins}, {Gonzalez-Perez}, {Lacey}  \&
  {Baugh}}{{Wilkins} et~al.}{2012}]{wilkins12a}
{Wilkins} S.~M.,  {Gonzalez-Perez} V.,  {Lacey} C.~G.,   {Baugh} C.~M.,  2012,
  \mn@doi [\mnras] {10.1111/j.1365-2966.2012.21344.x}, \href
  {http://adsabs.harvard.edu/abs/2012MNRAS.424.1522W} {424, 1522}

\bibitem[\protect\citeauthoryear{{Winters}, {Fleischer}, {Le Bertre}  \&
  {Sedlmayr}}{{Winters} et~al.}{1997}]{winters97a}
{Winters} J.~M.,  {Fleischer} A.~J.,  {Le Bertre} T.,   {Sedlmayr} E.,  1997,
  \aap, \href {https://ui.adsabs.harvard.edu/abs/1997A&A...326..305W} {326,
  305}

\bibitem[\protect\citeauthoryear{{Witstok} et~al.,}{{Witstok}
  et~al.}{2023}]{witstok23a}
{Witstok} J.,  et~al., 2023, \mn@doi [\nat] {10.1038/s41586-023-06413-w}, \href
  {https://ui.adsabs.harvard.edu/abs/2023Natur.621..267W} {621, 267}

\bibitem[\protect\citeauthoryear{{Wolfire}, {McKee}, {Hollenbach}  \&
  {Tielens}}{{Wolfire} et~al.}{2003}]{wolfire03a}
{Wolfire} M.~G.,  {McKee} C.~F.,  {Hollenbach} D.,   {Tielens} A.~G.~G.~M.,
  2003, \mn@doi [\apj] {10.1086/368016}, \href
  {http://adsabs.harvard.edu/abs/2003ApJ...587..278W} {587, 278}

\bibitem[\protect\citeauthoryear{{Woosley} \& {Weaver}}{{Woosley} \&
  {Weaver}}{1995}]{woosley95a}
{Woosley} S.~E.,  {Weaver} T.~A.,  1995, \mn@doi [\apjs] {10.1086/192237},
  \href {https://ui.adsabs.harvard.edu/abs/1995ApJS..101..181W} {101, 181}

\bibitem[\protect\citeauthoryear{{Yasuda} \& {Kozasa}}{{Yasuda} \&
  {Kozasa}}{2012}]{yasuda12a}
{Yasuda} Y.,  {Kozasa} T.,  2012, \mn@doi [\apj] {10.1088/0004-637X/745/2/159},
  \href {https://ui.adsabs.harvard.edu/abs/2012ApJ...745..159Y} {745, 159}

\bibitem[\protect\citeauthoryear{{Zhang} et~al.,}{{Zhang}
  et~al.}{2024}]{zhang24a}
{Zhang} E.,  et~al., 2024, \mn@doi [arXiv/2406.10338]
  {10.48550/arXiv.2406.10338}, \href
  {https://ui.adsabs.harvard.edu/abs/2024arXiv240610338Z} {p. arXiv:2406.10338}

\bibitem[\protect\citeauthoryear{{Zhao} \& {Furlanetto}}{{Zhao} \&
  {Furlanetto}}{2024}]{zhao24a}
{Zhao} J.,  {Furlanetto} S.~R.,  2024, \mn@doi [arXiv/2401.07893]
  {10.48550/arXiv.2401.07893}, \href
  {https://ui.adsabs.harvard.edu/abs/2024arXiv240107893Z} {p. arXiv:2401.07893}

\bibitem[\protect\citeauthoryear{{Zhukovska}, {Dobbs}, {Jenkins}  \&
  {Klessen}}{{Zhukovska} et~al.}{2016}]{zhukovska16a}
{Zhukovska} S.,  {Dobbs} C.,  {Jenkins} E.~B.,   {Klessen} R.~S.,  2016,
  \mn@doi [\apj] {10.3847/0004-637X/831/2/147}, \href
  {https://ui.adsabs.harvard.edu/abs/2016ApJ...831..147Z} {831, 147}

\bibitem[\protect\citeauthoryear{{Ziparo}, {Ferrara}, {Sommovigo}  \&
  {Kohandel}}{{Ziparo} et~al.}{2023}]{ziparo23a}
{Ziparo} F.,  {Ferrara} A.,  {Sommovigo} L.,   {Kohandel} M.,  2023, \mn@doi
  [\mnras] {10.1093/mnras/stad125}, \href
  {https://ui.adsabs.harvard.edu/abs/2023MNRAS.520.2445Z} {520, 2445}

\makeatother
\end{thebibliography}
\input{ms.bbl}

\section{Appendix}
\subsection{The Impact of the Initial Grain Size Distribution on Dust Masses.}
\begin{figure}
  \includegraphics[scale=0.4]{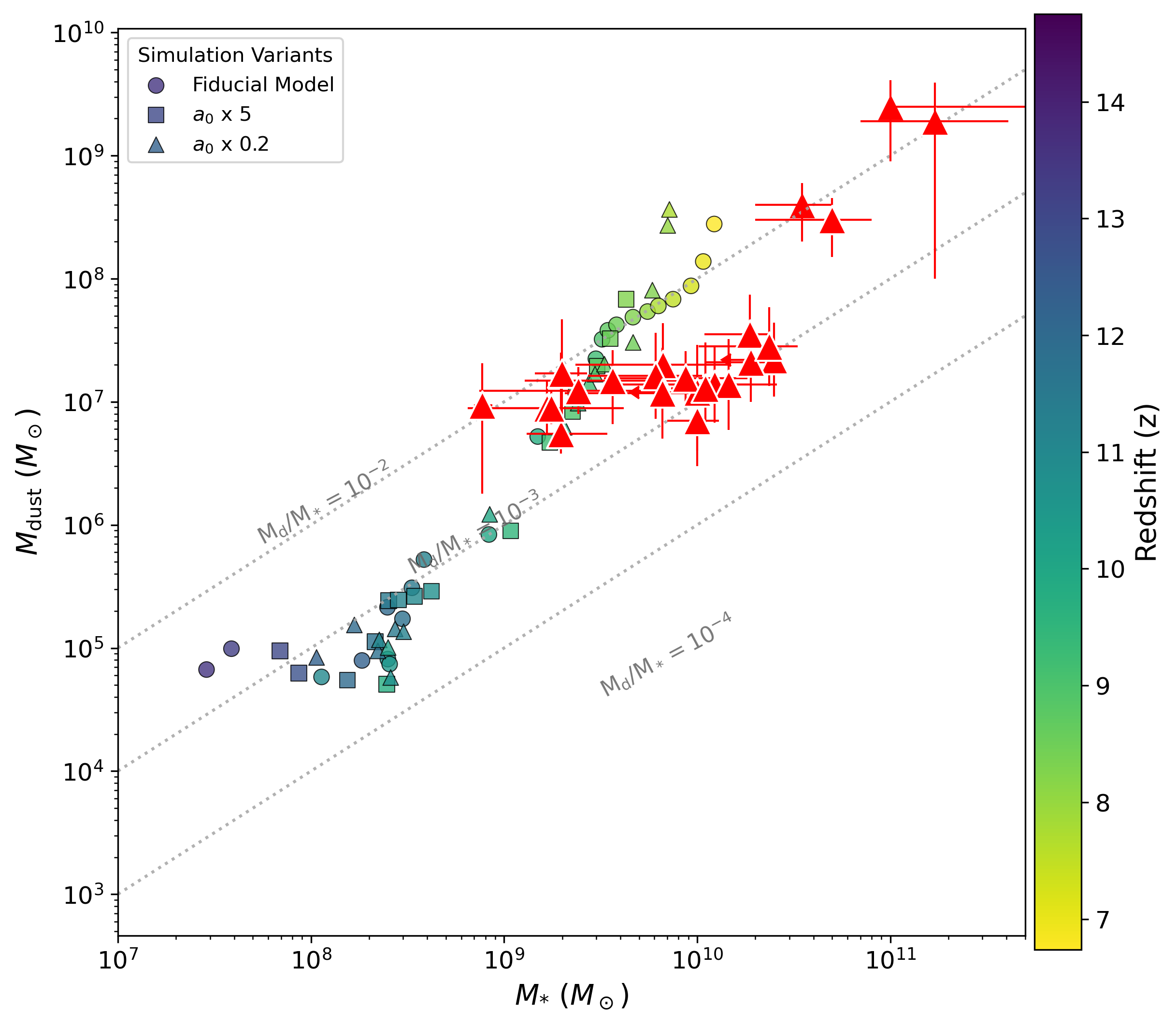}
  \caption{{\bf Demonstration of the relative lack of impact of our assumed dust size distribution upon initialization on our final results.}  We show the results of three model runs for an example galaxy: a fiducial run, one in which we've reduced the median in the initialized grain size distribution by a factor $5$, and one in which we've increased the median by a factor $5$.  The impact on the galaxy dust masses (as shown via the redshift evolution of the $M_{\
  rm dust}-M_*$ relation) is minimal. \label{figure:appendix1}}
\end{figure}

As discussed in \S~\ref{section:methods}, we initialize the grain size distributions as lognormals, with the initial size and width dependent on the source of dust (i.e., SNe or AGB).   In this Appendix, we demonstrate the relative lack of impact this choice makes on our final dust masses.  In Figure~\ref{figure:appendix1}, we show the results of three model runs of an example galaxy: our fiducial setup, one in which the median size in the lognormal is reduced by a factor 5, and one in which it is increased by a factor 5.   We show the redshift evolution of these model galaxies in $M_{\rm dust}-M_*$ space.  The three runs have extremely similar results, owing to the fact that interstellar processes rapidly erase the memory of the initialization of the dust grain size distribution. 

\subsection{Maximum Dust-to-Stellar Mass Ratio from Stellar Production}
\label{section:appendix_yields}
In this section, we compute the expected maximum  $M_{\rm
  dust}/M_*$ that can be expected from our adopted stellar yields and
condensation efficiencies alone, in the absence of any ISM growth or
destruction.  For a stellar population formed with a given initial mass function (IMF), the
maximum dust-to-stellar mass ratio from production alone is:
\begin{equation}
\label{equation:yields}
\left(\frac{M_{\rm dust}}{M_*}\right)_{\rm prod} = \frac{\sum_i \int_{m_{\rm low,i}}^{m_{\rm up}} \delta_i \cdot y_i(m) \cdot \phi(m) \, dm}{\int_{m_{\rm low,i}}^{m_{\rm up}} m \cdot \phi(m)\, dm}
\end{equation}
where $\phi(m)$ is the adopted IMF, $y_i(m)$ is the net
yield (mass of newly synthesized element $i$ returned to the ISM) from
a star of initial mass $m$, and $\delta_i$ is the condensation
efficiency for element $i$.  The sum is over all condensable species, and the integrals run over SN progenitors ($m \ge
8 M_\odot$) and AGB stars ($m \le 8 M_\odot$) separately.

We perform these integrals numerically for a fixed assumed low
  metallicity ($Z/Z_{\odot} \approx 10^{-2}$), and find:
\begin{equation}
\left(\frac{M_{\rm dust}}{M_*}\right)_{\rm prod}^{\rm max} \approx 3.4 \times 10^{-3}
\end{equation}
which is in good agreement with the analytic model of
  \citet{ferrara25a}. It is important to note that our numerical
  calculation is a theoretical upper limit from our simulations, and
  does not include the impact of destruction processes near the sites
  of star formation, which will reduce the effective dust-to-stellar
  ratio in the absence of growth (c.f. Figure~\ref{figure:dust_grow}).

\subsection{Reliability of Machine Learning Model}  
\begin{figure}
  \includegraphics[scale=0.4]{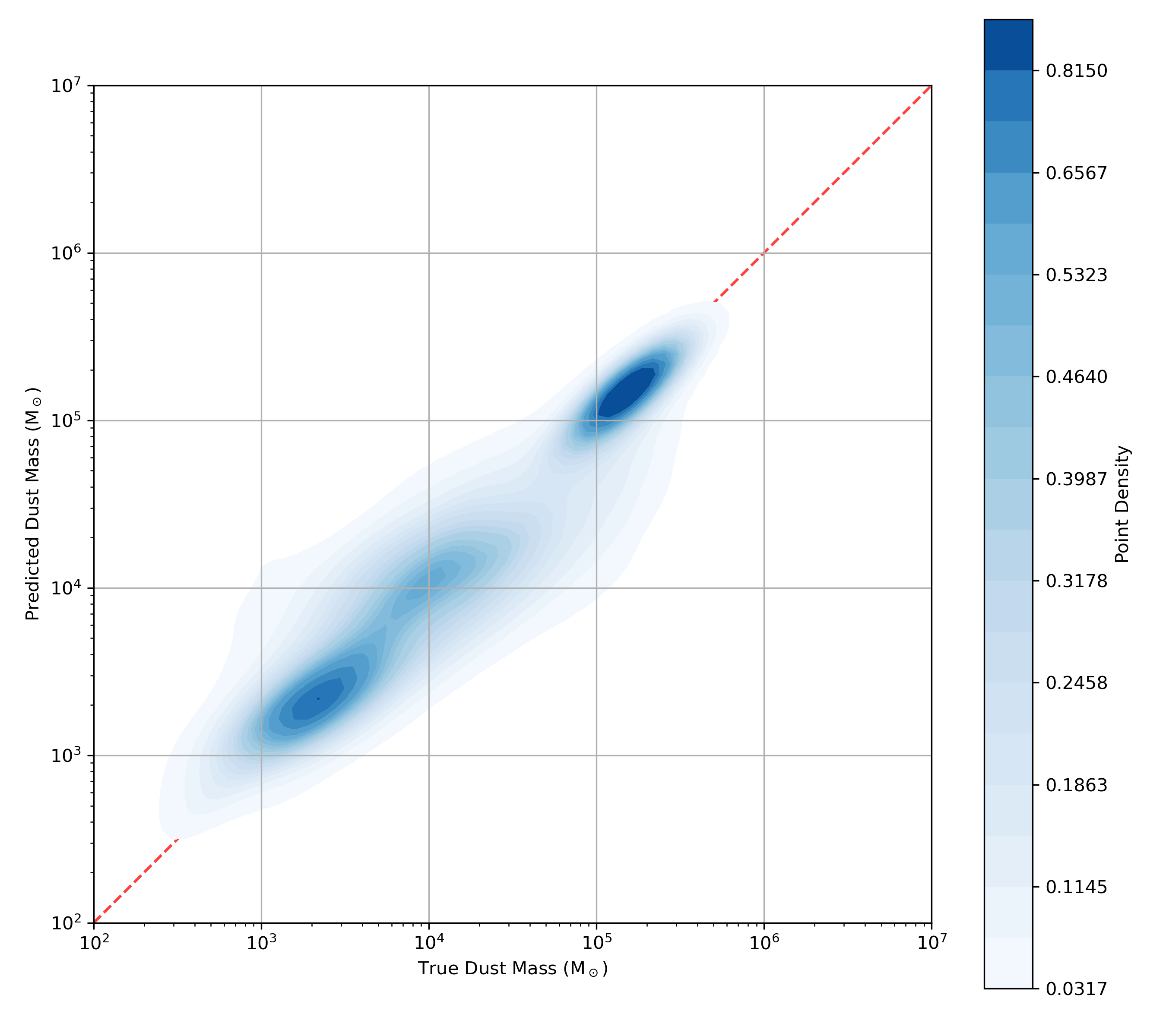}
  \caption{{\bf Demonstration of the reliability of our trained machine learning model on the dust particles in our simulation.}  Points represent individual dust particles in our example galaxy, and have been split into an 80/20 training/testing set.  We apply our ML model to the $20\%$ testing set, and show the predicted vs true dust masses here.  Points close to the 1:1 line demonstrate accurate predictions from our model.\label{figure:appendix2}}
\end{figure}

In Figure~\ref{figure:appendix2}, we demonstrate the reliability of our trained {\sc xgboost} model on predicting the dust masses from the individual physical properties of the dust particles.  To do this, we employ an 80/20 split between trained/sequestered model results, and plot the predicted vs true dust masses for our example galaxy.  As is clear, the majority of the points follow the 1:1 line, demonstrating reasonable model reliability.

\end{document}